\documentstyle[12pt,epsf]{article}

\makeatletter
\newcount\GRAPHICSTYPE
\GRAPHICSTYPE=\z@
\def\GRAPHICSPS#1{%
 \ifcase\GRAPHICSTYPE
  ps: #1%
 \or
  language "PS", include "#1"%
 \fi
}%
\def\graffile#1#2#3#4{%
    \leavevmode
    \raise -#4 \BOXTHEFRAME{%
        \hbox to #2{\raise #3\hbox{\null #1}}}%
}%
\def\draftbox#1#2#3#4{%
 \leavevmode\raise -#4 \hbox{%
  \frame{\rlap{\protect\tiny #1}\hbox to #2%
   {\vrule height#3 width\z@ depth\z@\hfil}%
  }%
 }%
}%
\newcount\draft
\draft=\z@
\def\GRAPHIC#1#2#3#4#5{%
 \ifnum\draft=\@ne\draftbox{#2}{#3}{#4}{#5}%
  \else\graffile{#1}{#3}{#4}{#5}%
  \fi
 }%
\def\addtoLaTeXparams#1{%
    \edef\LaTeXparams{\LaTeXparams #1}}%
\newif\ifBoxFrame \BoxFramefalse
\newif\ifOverFrame \OverFramefalse

\def\BOXTHEFRAME#1{%
   \hbox{%
      \ifBoxFrame
         \frame{#1}%
      \else
         {#1}%
      \fi
   }%
}

\def\doFRAMEparams#1{\BoxFramefalse\OverFramefalse\readFRAMEparams#1\end}%
\def\readFRAMEparams#1{%
 \ifx#1\end%
  \let\next=\relax
  \else
  \ifx#1i\dispkind=\z@\fi
  \ifx#1d\dispkind=\@ne\fi
  \ifx#1f\dispkind=\tw@\fi
  \ifx#1t\addtoLaTeXparams{t}\fi
  \ifx#1b\addtoLaTeXparams{b}\fi
  \ifx#1p\addtoLaTeXparams{p}\fi
  \ifx#1h\addtoLaTeXparams{h}\fi
  \ifx#1X\BoxFrametrue\fi
  \ifx#1O\OverFrametrue\fi
  \let\next=\readFRAMEparams
  \fi
 \next
 }%
\def\IFRAME#1#2#3#4#5#6{%
      \bgroup
      \parindent=0pt%
      \setbox0 = \hbox{#6}%
      \@tempdima = #1%
      \ifOverFrame
          \typeout{This is not implemented yet}%
          \show\HELP
      \else
         \ifdim\wd0>\@tempdima
            \advance\@tempdima by \@tempdima
            \ifdim\wd0 >\@tempdima
               \textwidth=\@tempdima
               \setbox1 =\vbox{%
                  \noindent\hbox to
\@tempdima{\hfill\GRAPHIC{#5}{#4}{#1}{#2}{#3}\hfill}\\%
                  \noindent\hbox to \@tempdima{\parbox[b]{\@tempdima}{#6}}%
               }%
               \wd1=\@tempdima
            \else
               \textwidth=\wd0
               \setbox1 =\vbox{%
                 \noindent\hbox to
\wd0{\hfill\GRAPHIC{#5}{#4}{#1}{#2}{#3}\hfill}\\%
                 \noindent\hbox{#6}%
               }%
               \wd1=\wd0
            \fi
         \else
            \hsize=\@tempdima
            \setbox1 =\vbox{%
                \unskip\GRAPHIC{#5}{#4}{#1}{#2}{0pt}%
                \break
                \unskip\hbox to \@tempdima{\hfill #6\hfill}%
            }%
            \wd1=\@tempdima
         \fi
         \@tempdimb=\ht1
         \advance\@tempdimb by \dp1
         \advance\@tempdimb by -#2%
         \advance\@tempdimb by #3%
         \leavevmode
         \raise -\@tempdimb \hbox{\box1}%
      \fi
      \egroup
}%
\def\DFRAME#1#2#3#4#5{%
 \begin{center}
     \ifOverFrame
        #5\par
     \fi
     \GRAPHIC{#4}{#3}{#1}{#2}{\z@}
     \ifOverFrame \else
        \par #5
     \fi
 \end{center}%
 }%
\def\FFRAME#1#2#3#4#5#6#7{%
 \begin{figure}[#1]%
  \begin{center}\GRAPHIC{#7}{#6}{#2}{#3}{\z@}\end{center}%
  \caption{\label{#5}#4}%
  \end{figure}%
 }%
\newcount\dispkind%
\def\FRAME#1#2#3#4#5#6#7#8{%
 \def\LaTeXparams{}%
 \dispkind=\z@
 \def\LaTeXparams{}%
 \doFRAMEparams{#1}%
 \ifnum\dispkind=\z@\IFRAME{#2}{#3}{#4}{#7}{#8}{#5}\else
  \ifnum\dispkind=\@ne\DFRAME{#2}{#3}{#7}{#8}{#5}\else
   \ifnum\dispkind=\tw@
    \edef\@tempa{\noexpand\FFRAME{\LaTeXparams}}%
    \@tempa{#2}{#3}{#5}{#6}{#7}{#8}%
    \fi
   \fi
  \fi
 }%
\def\TEXUX#1{"texux"}
\def\@@eqncr{\let\@tempa\relax
    \ifcase\@eqcnt \def\@tempa{& & &}\or \def\@tempa{& &}%
      \else \def\@tempa{&}\fi
     \@tempa
     \if@eqnsw
        \iftag@
           \@taggnum
        \else
           \@eqnnum\stepcounter{equation}\fi
     \fi
     \global\tag@false
     \global\@eqnswtrue
     \global\@eqcnt\z@\cr}

 \newif\iftag@ \tag@false
 \def\tag{\@ifnextchar*{\@tagstar}{\@tag}}
 \def\@tag#1{%
     \global\tag@true
     \global\def\@taggnum{(#1)}}
 \def\@tagstar*#1{%
     \global\tag@true
     \global\def\@taggnum{#1}%
}
\long\def\QQQ#1#2{%
     \long\expandafter\def\csname#1\endcsname{#2}}%
\@ifundefined{QTP}{\def\QTP#1{}}{}
\@ifundefined{Qcb}{}{}
\@ifundefined{Qct}{}{}
\@ifundefined{Qlb}{}{}
\@ifundefined{Qlt}{}{}
\long\def\QQA#1#2{}%
\def\QTR#1#2{{\csname#1\endcsname #2}}
\def\EXPAND#1[#2]#3{}%
\def\NOEXPAND#1[#2]#3{}%
\def\LaTeXparent#1{}%
\def\ChildStyles#1{}%
\def\ChildDefaults#1{}%
\def\QTagDef#1#2#3{}%
\def\QQfnmark#1{\footnotemark}

\@ifundefined{INDEX}{\def\INDEX#1#2{}{}}{}%
\@ifundefined{SUBINDEX}{\def\SUBINDEX#1#2#3{}{}{}}{}%
\def\initial#1{\bigbreak{\raggedright\large\bf #1}\kern 2\p@
   \penalty3000}%
\def\entry#1#2{\item {#1}, #2}%
\@ifundefined{ZZZ}{}{\makeatletter\input gnuindex.sty\makeatother\makeindex\makeatletter}%
\@ifundefined{abstract}{%
 \def\abstract{%
  \if@twocolumn
   \section*{Abstract (Not appropriate in this style!)}%
   \else \small
   \begin{center}{\bf Abstract\vspace{-.5em}\vspace{\z@}}\end{center}%
   \quotation
   \fi
  }%
 }{%
 }%
\@ifundefined{endabstract}{\def\endabstract
  {\if@twocolumn\else\endquotation\fi}}{}%
\@ifundefined{maketitle}{\def\maketitle#1{}}{}%
\@ifundefined{affiliation}{\def\affiliation#1{}}{}%
\@ifundefined{proof}{}{}%
\@ifundefined{endproof}{}{}%
\@ifundefined{newfield}{\def\newfield#1#2{}}{}%
\@ifundefined{chapter}{\def\chapter#1{\par(Chapter head:)#1\par }%
 \newcount\c@chapter}{}%
\@ifundefined{part}{\def\part#1{\par(Part head:)#1\par }}{}%
\@ifundefined{section}{\def\section#1{\par(Section head:)#1\par }}{}%
\@ifundefined{subsection}{\def\subsection#1%
 {\par(Subsection head:)#1\par }}{}%
\@ifundefined{subsubsection}{\def\subsubsection#1%
 {\par(Subsubsection head:)#1\par }}{}%
\@ifundefined{paragraph}{\def\paragraph#1%
 {\par(Subsubsubsection head:)#1\par }}{}%
\@ifundefined{subparagraph}{\def\subparagraph#1%
 {\par(Subsubsubsubsection head:)#1\par }}{}%
\@ifundefined{therefore}{}{}%
\@ifundefined{backepsilon}{}{}%
\@ifundefined{yen}{}{}%
\@ifundefined{registered}{%
   \def\registered{\relax\ifmmode{}\r@gistered
                    \else$\m@th\r@gistered$\fi}%
 \def\r@gistered{^{\ooalign
  {\hfil\raise.07ex\hbox{$\scriptstyle\rm\text{R}$}\hfil\crcr
  \mathhexbox20D}}}}{}%
\@ifundefined{Eth}{}{}%
\@ifundefined{eth}{}{}%
\@ifundefined{Thorn}{}{}%
\@ifundefined{thorn}{}{}%
\@ifundefined{degree}{}{}%
\def\BibTeX{{\rm B\kern-.05em{\sc i\kern-.025em b}\kern-.08em
    T\kern-.1667em\lower.7ex\hbox{E}\kern-.125emX}}%
\newdimen\theight
\def\Column{%
 \vadjust{\setbox\z@=\hbox{\scriptsize\quad\quad tcol}%
  \theight=\ht\z@\advance\theight by \dp\z@\advance\theight by \lineskip
  \kern -\theight \vbox to \theight{%
   \rightline{\rlap{\box\z@}}%
   \vss
   }%
  }%
 }%
\def\qed{%
 \ifhmode\unskip\nobreak\fi\ifmmode\ifinner\else\hskip5\p@\fi\fi
 \hbox{\hskip5\p@\vrule width4\p@ height6\p@ depth1.5\p@\hskip\p@}%
 }%
\def\miss{\hbox{\vrule height2\p@ width 2\p@ depth\z@}}%
\def\tcol#1{{\baselineskip=6\p@ \vcenter{#1}} \Column}  %
\def\newfmtname{LaTeX2e}
\def\chkcompat{%
   \if@compatibility
   \else
     \usepackage{latexsym}
   \fi
}

\ifx\fmtname\newfmtname
  \DeclareOldFontCommand{\rm}{\normalfont\rmfamily}{\mathrm}
  \DeclareOldFontCommand{\sf}{\normalfont\sffamily}{\mathsf}
  \DeclareOldFontCommand{\tt}{\normalfont\ttfamily}{\mathtt}
  \DeclareOldFontCommand{\bf}{\normalfont\bfseries}{\mathbf}
  \DeclareOldFontCommand{\it}{\normalfont\itshape}{\mathit}
  \DeclareOldFontCommand{\sl}{\normalfont\slshape}{\@nomath\sl}
  \DeclareOldFontCommand{\sc}{\normalfont\scshape}{\@nomath\sc}
  \chkcompat
\fi

\def\alpha{\Greekmath 010B }%
\def\beta{\Greekmath 010C }%
\def\gamma{\Greekmath 010D }%
\def\delta{\Greekmath 010E }%
\def\epsilon{\Greekmath 010F }%
\def\zeta{\Greekmath 0110 }%
\def\eta{\Greekmath 0111 }%
\def\theta{\Greekmath 0112 }%
\def\iota{\Greekmath 0113 }%
\def\kappa{\Greekmath 0114 }%
\def\lambda{\Greekmath 0115 }%
\def\mu{\Greekmath 0116 }%
\def\nu{\Greekmath 0117 }%
\def\xi{\Greekmath 0118 }%
\def\pi{\Greekmath 0119 }%
\def\rho{\Greekmath 011A }%
\def\sigma{\Greekmath 011B }%
\def\tau{\Greekmath 011C }%
\def\upsilon{\Greekmath 011D }%
\def\phi{\Greekmath 011E }%
\def\chi{\Greekmath 011F }%
\def\psi{\Greekmath 0120 }%
\def\omega{\Greekmath 0121 }%
\def\varepsilon{\Greekmath 0122 }%
\def\vartheta{\Greekmath 0123 }%
\def\varpi{\Greekmath 0124 }%
\def\varrho{\Greekmath 0125 }%
\def\varsigma{\Greekmath 0126 }%
\def\varphi{\Greekmath 0127 }%

\def\nabla{\Greekmath 0272}

\def\GreekBold{\@ne}%
\def\One{\@ne}

\def\Greekmath#1#2#3#4{%
    \ifx\GreekBold\One
        \mathchar"#1#2#3#4%
    \else
		\mbox{\boldmath$\mathchar"#1#2#3#4$}
	\fi}

\let\SAVEPBF=\pbf

\def\pbf{\let\GreekBold = \relax\SAVEPBF}%

\expandafter\ifx\csname ds@amstex\endcsname\relax
\else\makeatother\endinput\fi

\let\DOTSI\relax
\def\RIfM@{\relax\ifmmode}%
\def\FN@{\futurelet\next}%
\newcount\intno@
\def\iint{\DOTSI\intno@\tw@\FN@\ints@}%
\def\iiint{\DOTSI\intno@\thr@@\FN@\ints@}%
\def\iiiint{\DOTSI\intno@4 \FN@\ints@}%
\def\idotsint{\DOTSI\intno@\z@\FN@\ints@}%
\def\ints@{\findlimits@\ints@@}%
\newif\iflimtoken@
\newif\iflimits@
\def\findlimits@{\limtoken@true\ifx\next\limits\limits@true
 \else\ifx\next\nolimits\limits@false\else
 \limtoken@false\ifx\ilimits@\nolimits\limits@false\else
 \ifinner\limits@false\else\limits@true\fi\fi\fi\fi}%
\def\multint@{\int\ifnum\intno@=\z@\intdots@                          
 \else\intkern@\fi                                                    
 \ifnum\intno@>\tw@\int\intkern@\fi                                   
 \ifnum\intno@>\thr@@\int\intkern@\fi                                 
 \int}
\def\multintlimits@{\intop\ifnum\intno@=\z@\intdots@\else\intkern@\fi
 \ifnum\intno@>\tw@\intop\intkern@\fi
 \ifnum\intno@>\thr@@\intop\intkern@\fi\intop}%
\def\intic@{%
    \mathchoice{\hskip.5em}{\hskip.4em}{\hskip.4em}{\hskip.4em}}%
\def\negintic@{\mathchoice
 {\hskip-.5em}{\hskip-.4em}{\hskip-.4em}{\hskip-.4em}}%
\def\ints@@{\iflimtoken@                                              
 \def\ints@@@{\iflimits@\negintic@
   \mathop{\intic@\multintlimits@}\limits                             
  \else\multint@\nolimits\fi                                          
  \eat@}
 \else                                                                
 \def\ints@@@{\iflimits@\negintic@
  \mathop{\intic@\multintlimits@}\limits\else
  \multint@\nolimits\fi}\fi\ints@@@}%
\def\intkern@{\mathchoice{\!\!\!}{\!\!}{\!\!}{\!\!}}%
\def\plaincdots@{\mathinner{\cdotp\cdotp\cdotp}}%
\def\intdots@{\mathchoice{\plaincdots@}%
 {{\cdotp}\mkern1.5mu{\cdotp}\mkern1.5mu{\cdotp}}%
 {{\cdotp}\mkern1mu{\cdotp}\mkern1mu{\cdotp}}%
 {{\cdotp}\mkern1mu{\cdotp}\mkern1mu{\cdotp}}}%
\def\RIfM@{\relax\protect\ifmmode}
\def\text{\RIfM@\expandafter\text@\else\expandafter\mbox\fi}
\let\nfss@text\text
\def\text@#1{\mathchoice
   {\textdef@\displaystyle\f@size{#1}}%
   {\textdef@\textstyle\tf@size{\firstchoice@false #1}}%
   {\textdef@\textstyle\sf@size{\firstchoice@false #1}}%
   {\textdef@\textstyle \ssf@size{\firstchoice@false #1}}%
   \glb@settings}

\def\textdef@#1#2#3{\hbox{{%
                    \everymath{#1}%
                    \let\f@size#2\selectfont
                    #3}}}
\newif\iffirstchoice@
\firstchoice@true
\def\Let@{\relax\iffalse{\fi\let\\=\cr\iffalse}\fi}%
\def\vspace@{\def\vspace##1{\crcr\noalign{\vskip##1\relax}}}%
\def\multilimits@{\bgroup\vspace@\Let@
 \baselineskip\fontdimen10 \scriptfont\tw@
 \advance\baselineskip\fontdimen12 \scriptfont\tw@
 \lineskip\thr@@\fontdimen8 \scriptfont\thr@@
 \lineskiplimit\lineskip
 \vbox\bgroup\ialign\bgroup\hfil$\m@th\scriptstyle{##}$\hfil\crcr}%
\def\Sb{_\multilimits@}%
\def\endSb{\crcr\egroup\egroup\egroup}%
\def\Sp{^\multilimits@}%

\newdimen\ex@
\def\rightarrowfill@#1{$#1\m@th\mathord-\mkern-6mu\cleaders
 \hbox{$#1\mkern-2mu\mathord-\mkern-2mu$}\hfill
 \mkern-6mu\mathord\rightarrow$}%
\def\leftarrowfill@#1{$#1\m@th\mathord\leftarrow\mkern-6mu\cleaders
 \hbox{$#1\mkern-2mu\mathord-\mkern-2mu$}\hfill\mkern-6mu\mathord-$}%
\def\leftrightarrowfill@#1{$#1\m@th\mathord\leftarrow
\mkern-6mu\cleaders
 \hbox{$#1\mkern-2mu\mathord-\mkern-2mu$}\hfill
 \mkern-6mu\mathord\rightarrow$}%
\def\overrightarrow{\mathpalette\overrightarrow@}%
\def\overrightarrow@#1#2{\vbox{\ialign{##\crcr\rightarrowfill@#1\crcr
 \noalign{\kern-\ex@\nointerlineskip}$\m@th\hfil#1#2\hfil$\crcr}}}%

\def\overleftarrow{\mathpalette\overleftarrow@}%
\def\overleftarrow@#1#2{\vbox{\ialign{##\crcr\leftarrowfill@#1\crcr
 \noalign{\kern-\ex@\nointerlineskip}$\m@th\hfil#1#2\hfil$\crcr}}}%
\def\overleftrightarrow{\mathpalette\overleftrightarrow@}%
\def\overleftrightarrow@#1#2{\vbox{\ialign{##\crcr
   \leftrightarrowfill@#1\crcr
 \noalign{\kern-\ex@\nointerlineskip}$\m@th\hfil#1#2\hfil$\crcr}}}%
\def\underrightarrow{\mathpalette\underrightarrow@}%
\def\underrightarrow@#1#2{\vtop{\ialign{##\crcr$\m@th\hfil#1#2\hfil
  $\crcr\noalign{\nointerlineskip}\rightarrowfill@#1\crcr}}}%

\def\underleftarrow{\mathpalette\underleftarrow@}%
\def\underleftarrow@#1#2{\vtop{\ialign{##\crcr$\m@th\hfil#1#2\hfil
  $\crcr\noalign{\nointerlineskip}\leftarrowfill@#1\crcr}}}%
\def\underleftrightarrow{\mathpalette\underleftrightarrow@}%
\def\underleftrightarrow@#1#2{\vtop{\ialign{##\crcr$\m@th
  \hfil#1#2\hfil$\crcr
 \noalign{\nointerlineskip}\leftrightarrowfill@#1\crcr}}}%

\def\qopnamewl@#1{\mathop{\operator@font#1}\nlimits@}
\let\nlimits@\displaylimits
\def\setboxz@h{\setbox\z@\hbox}

\def\varlim@#1#2{\mathop{\vtop{\ialign{##\crcr
 \hfil$#1\m@th\operator@font lim$\hfil\crcr
 \noalign{\nointerlineskip}#2#1\crcr
 \noalign{\nointerlineskip\kern-\ex@}\crcr}}}}

 \def\rightarrowfill@#1{\m@th\setboxz@h{$#1-$}\ht\z@\z@
  $#1\copy\z@\mkern-6mu\cleaders
  \hbox{$#1\mkern-2mu\box\z@\mkern-2mu$}\hfill
  \mkern-6mu\mathord\rightarrow$}
\def\leftarrowfill@#1{\m@th\setboxz@h{$#1-$}\ht\z@\z@
  $#1\mathord\leftarrow\mkern-6mu\cleaders
  \hbox{$#1\mkern-2mu\copy\z@\mkern-2mu$}\hfill
  \mkern-6mu\box\z@$}

\def\projlim{\qopnamewl@{proj\,lim}}
\def\injlim{\qopnamewl@{inj\,lim}}
\def\varinjlim{\mathpalette\varlim@\rightarrowfill@}
\def\varprojlim{\mathpalette\varlim@\leftarrowfill@}
\def\varliminf{\mathpalette\varliminf@{}}
\def\varliminf@#1{\mathop{\underline{\vrule\@depth.2\ex@\@width\z@
   \hbox{$#1\m@th\operator@font lim$}}}}
\def\varlimsup{\mathpalette\varlimsup@{}}
\def\varlimsup@#1{\mathop{\overline
  {\hbox{$#1\m@th\operator@font lim$}}}}

\begingroup \catcode `|=0 \catcode `[= 1
\catcode`]=2 \catcode `\{=12 \catcode `\}=12
\catcode`\\=12
|gdef|@alignverbatim#1\end{align}[#1|end[align]]
|gdef|@salignverbatim#1\end{align*}[#1|end[align*]]

|gdef|@alignatverbatim#1\end{alignat}[#1|end[alignat]]
|gdef|@salignatverbatim#1\end{alignat*}[#1|end[alignat*]]

|gdef|@xalignatverbatim#1\end{xalignat}[#1|end[xalignat]]
|gdef|@sxalignatverbatim#1\end{xalignat*}[#1|end[xalignat*]]

|gdef|@gatherverbatim#1\end{gather}[#1|end[gather]]
|gdef|@sgatherverbatim#1\end{gather*}[#1|end[gather*]]

|gdef|@gatherverbatim#1\end{gather}[#1|end[gather]]
|gdef|@sgatherverbatim#1\end{gather*}[#1|end[gather*]]

|gdef|@multilineverbatim#1\end{multiline}[#1|end[multiline]]
|gdef|@smultilineverbatim#1\end{multiline*}[#1|end[multiline*]]

|gdef|@arraxverbatim#1\end{arrax}[#1|end[arrax]]
|gdef|@sarraxverbatim#1\end{arrax*}[#1|end[arrax*]]

|gdef|@tabulaxverbatim#1\end{tabulax}[#1|end[tabulax]]
|gdef|@stabulaxverbatim#1\end{tabulax*}[#1|end[tabulax*]]

|endgroup

\def\align{\@verbatim \frenchspacing\@vobeyspaces \@alignverbatim
You are using the "align" environment in a style in which it is not defined.}

\@namedef{align*}{\@verbatim\@salignverbatim
You are using the "align*" environment in a style in which it is not defined.}
\expandafter\let\csname endalign*\endcsname =\endtrivlist

\def\alignat{\@verbatim \frenchspacing\@vobeyspaces \@alignatverbatim
You are using the "alignat" environment in a style in which it is not defined.}

\@namedef{alignat*}{\@verbatim\@salignatverbatim
You are using the "alignat*" environment in a style in which it is not
defined.}
\expandafter\let\csname endalignat*\endcsname =\endtrivlist

\def\xalignat{\@verbatim \frenchspacing\@vobeyspaces \@xalignatverbatim
You are using the "xalignat" environment in a style in which it is not
defined.}

\@namedef{xalignat*}{\@verbatim\@sxalignatverbatim
You are using the "xalignat*" environment in a style in which it is not
defined.}
\expandafter\let\csname endxalignat*\endcsname =\endtrivlist

\def\gather{\@verbatim \frenchspacing\@vobeyspaces \@gatherverbatim
You are using the "gather" environment in a style in which it is not defined.}

\@namedef{gather*}{\@verbatim\@sgatherverbatim
You are using the "gather*" environment in a style in which it is not defined.}
\expandafter\let\csname endgather*\endcsname =\endtrivlist

\def\multiline{\@verbatim \frenchspacing\@vobeyspaces \@multilineverbatim
You are using the "multiline" environment in a style in which it is not
defined.}

\@namedef{multiline*}{\@verbatim\@smultilineverbatim
You are using the "multiline*" environment in a style in which it is not
defined.}
\expandafter\let\csname endmultiline*\endcsname =\endtrivlist

\def\arrax{\@verbatim \frenchspacing\@vobeyspaces \@arraxverbatim
You are using a type of "array" construct that is only allowed in AmS-LaTeX.}

\def\tabulax{\@verbatim \frenchspacing\@vobeyspaces \@tabulaxverbatim
You are using a type of "tabular" construct that is only allowed in AmS-LaTeX.}

\@namedef{arrax*}{\@verbatim\@sarraxverbatim
You are using a type of "array*" construct that is only allowed in AmS-LaTeX.}
\expandafter\let\csname endarrax*\endcsname =\endtrivlist

\@namedef{tabulax*}{\@verbatim\@stabulaxverbatim
You are using a type of "tabular*" construct that is only allowed in
AmS-LaTeX.}
\expandafter\let\csname endtabulax*\endcsname =\endtrivlist

\@ifundefined{theorem}{}{}
\@ifundefined{lemma}{}{}
\@ifundefined{corollary}{}{}
\@ifundefined{conjecture}{}{}
\@ifundefined{proposition}{}{}
\@ifundefined{axiom}{}{}
\@ifundefined{remark}{}{}
\@ifundefined{example}{}{}
\@ifundefined{exercise}{}{}
\@ifundefined{definition}{}{}

\makeatother

    \def\initial#1{\bigbreak{\raggedright\large\bf #1}\kern 2pt\penalty3000}
    \def\entry#1#2{\item {#1}, #2}

  \def\INDEX{\@bsphack\begingroup\@sanitize\@WRINDEX\@indexfile}
  \def\@WRINDEX#1#2#3{\let\thepage\relax
     \xdef\@gtempa{\write#1{\string
      \entry{#2}{\thepage}{#3}}}\endgroup\@gtempa
     \if@nobreak \ifvmode\nobreak\fi\fi\@esphack}
  \def\SUBINDEX{\@bsphack\begingroup\@sanitize\@WRSUBINDEX\@indexfile}
  \def\@WRSUBINDEX#1#2#3#4{\let\thepage\relax
     \xdef\@gtempa{\write#1{\string
     \entry{#2\space\space\space#4}{\thepage}{#3}{#4}}}\endgroup\@gtempa
     \if@nobreak \ifvmode\nobreak\fi\fi\@esphack}

\QQQ{Language}{
American English
}

\begin{document}

\title{\noindent {\small{hep-th/9411217 \hfill
USC-94/HEP-B3}}\bigskip\bigskip \\
Classical Solutions of 2D String Theory\\
in any Curved Spacetime\thanks{%
Based on lectures delivered at the Journ\'ee Cosmologique, Observatoire de
Paris, May 1994. To appear in the proceedings.}}
\author{Itzhak Bars\thanks{%
Research supported in part by the Deparment of Energy under Grant No.
DE-FG03-84ER40168.} \\
Department of Physics and Astronomy\\
University of Southern California\\
Los Angeles, CA 90089-0484}
\date{}
\maketitle

\begin{abstract}
The complete set of solutions of two dimensional classical string theory are
constructed for any curved spacetime. They describe folded strings moving in
curved spacetime. Surprizing stringy behavior becomes evident at
singularities such as black holes. The solutions are given in the form of a
map from the world sheet to target spacetime, where the world sheet has to
be divided into lattice-like patches corresponding to different maps. A
recursion relation analogous to a ``transfer matrix'' that connects these
maps into a single continuous map is derived. This ``transfer matrix''
encodes the properties of the world sheet lattice on the one hand and the
geometry of spacetime on the other hand. The solutions are completely
classified by their behavior in the asymptotically flat region of spacetime
where they reduce, as boundary conditions, to the folded string solutions
that have been known for 19 years.


\newpage\
\end{abstract}

\section{Introduction}

The original physical motivations for studying string theory were: (1)
understanding unification of forces including quantum gravity, and (2)
understanding the Standard Model. In recent years it has become more and
more evident that these goals should be examined in the presence of curved
4D space-time string backgrounds. The construction of 4D curved space-time
string theories that correspond to exact conformal theories have provided
models in which various questions can be investigated\cite{ibcurved}.

The usual scenario of flat 4D plus extra curved dimensions may not be the
right approach for making predictions about the Standard Model. I believe
that the gauge symmetries and spectrum of quark + lepton families, which are
the main ingredients of the Standard Model, were probably fixed during the
early times in the evolution of the Universe. At such times 4D space-time
was curved. Since curvature contributes to the central charge and other
topological aspects of String Theory, it is likely that the predictions of
String Theory under such conditions may be quite different than the flat 4D
approach. Therefore, I believe that String Theory in curved space-time must
be better understood before attempting to make connections to low energy
physics. One should consider all kinds of curved backgrounds, not only the
traditional cosmological backgrounds, since the passage from curved
space-time to flat space-time may involve various phase transitions,
including inflation of a small region of the original curved universe to
today's universe that is essentially homogeneous and flat. The gauge bosons,
and chiral families of quarks and leptons in a small region of the early
curved universe would become the ones observed in today's inflated flat
universe. The possibility of such a scenario suggests that curved space-time
string theory deserves intensive study. In addition, the issues surrounding
gravitational singularities should be answered in the context of curved
space-time string theory, as it is the only known theory of quantum gravity.

In this paper we are interested in two aspects of string theory: (i) strings
in curved space-time and (ii) folded strings. They are both explored
simultaneously in the complete classical solution of 2D string theory that
we present here. First, we feel it is important to understand classical
string theory in curved space-time in order to understand and interpret
quantum string theory in curved space-time. This is relevant to fundamental
questions of singularities in gravitational physics, as well as stringy
questions about the early universe and its influence on the low energy
spectrum of quarks and leptons. In 2D the only non-trivial stringy solutions
turn out to be necessarily folded strings, and therefore they are the only
path toward analyzing such questions in a toy model. Second, there has been
a long-standing interest in exploring consistent generalizations of
non-critical strings with the hope that they may be relevant for some branch
of physics. Folded strings fall into this category, especially in the area
of string-QCD relations.

Two dimensional string theory in {\it flat space-time} was discussed in 1975
by BBHP \cite{bbhp1} in the conformal, lightcone, and temporal gauges. The
classical folded string solutions were obtained, a semiclassical
quantization was performed, and agreement of results in various gauges was
displayed. Furthermore, the Lorentz covariance and consistency of the
quantum theory was proven by showing the closure of the 2D Poincar\'e
algebra, for which the proof was clearest in the lightcone gauge \cite{bbhp2}%
. BBHP discovered that the non-trivial classical motions correspond to
longitudinal oscillations of folded strings. Folds as well as end points (if
the string is open) move at the speed of light and oscillate against each
other. These massless points present an anomaly that needs to be treated
carefully. BBHP showed that by making these points massive, analyzing the
motion, and sending the mass to zero at the end, the physics could be
understood most satisfactorily. However, the mathematics is simplest in the
conformal gauge in the massless limit, where the same classical motions are
recovered provided one is careful \cite{bbhp1}.

The conformal gauge approach was recently generalized to curved space-time.
The general {\it classical folded string solutions} were obtained for any 2D
curved space-time and the motion in the 2D black hole was physically
interpreted \cite{ibjs}. It was shown that the admissible solutions are
those that smoothly connect to the folded string solutions in flat
space-time, since far away from singularities curved space-time approaches
flat space-time. Furthermore, by regarding curved space-time as a continuous
deformation of flat space-time, except for singularities, one can
intuitively guess the general behavior of the string motion as being similar
to the folded string solutions. Therefore, away from singularities, the
minimal 2D target space surface swept by the string turns out to be similar
to the one in flat space-time, but its detailed shape obeys certain global
conservation rules dictated by the curved metric and an associated
``transfer matrix''. In the vicinity of singularities the same conservation
laws are obeyed and they lead to certain surprises in the motion of strings.
Using these solutions the swallowing of a string by a black hole was
discussed, showing that new unusual stringy features emerge, such as the
tunneling of the string into the region beyond the black hole (the bare
singularity region) that is forbidden for particle geodesics, as well as
other new effects.

To avoid confusion it is useful to emphasize that the folded string states
are properties of the 2D quantum string theory as well. In a covariant
quantization, they exist in addition to the ``special momentum states'' of
the 2D quantum string theory that have been discussed in recent years. As
pointed out many times in our past work, folded 2D-string states are present
in the $d=2$ and $c\leq 25$ sector of the quantum theory. In simple string
models, when it has been possible to compute the spectrum, their norm is
positive and is proportional to $(c-26)$. Only if $d=2$ and $c=26$
simultaneously (e.g. $d=2$ flat space-time with linear dilaton such that $%
c=26$) the folded string states become zero norm states and then the special
discrete momentum states survive as the only stringy states. A simple model
in which these properties may be easily seen is the {\it covariant}
quantization of the 2D string theory, in which the physical states are
identified as the subset that satisfies the Virasoro constraints. For
example, it has been known for a long time that the $d\leq 25$ sector of the
flat string theory has non-trivial positive norm states (including for $d=2)$
that satisfy the Virasoro constraints and that there are no ghosts \cite
{noghost}.

The possibility of a more general{\it \ interacting }string theory that
includes folded strings, and the probable close connection with large-N QCD,
provides additional motivation for studying folded strings. For example, it
has been expected for a long time that there is an interacting string
version of QCD in 2D to 4D. Some relations that were discovered a long time
ago can be re-examined with a new point of view and generalized to curved
space-time. There are relations that involve zero fold strings as well as
folded strings:

\begin{itemize}
\item  The open, {\it zero fold} string, with spinors attached at the ends
\cite{barshanson}, and with interactions at the end points \cite
{ibequivalence}, was shown to reproduce the low orders of the perturbation
series of 2D large-N QCD with quarks \cite{ibequivalence}, in flat
space-time. This equivalence includes the 't Hooft spectrum \cite{thooft},
the 1/N strong interactions within QCD \cite{ccg}, and the 1/N Electro-weak
interactions with external fields \cite{einhorn}. These features were
reproduced within string theory by a many-body type action describing
propagation, string-string interactions at end points, and interactions with
external fields \cite{ibequivalence}.

\item  More recently it has become apparent \cite{ibberkeley} that folded
strings in flat 2D space-time are closely connected to the flat space-time
large-N 2D gauge theory interacting with fermions or bosons in the adjoint
representation \cite{klebanov} \cite{kutasov}. The QCD flux (i.e. string)
folds at the location of the adjoint fermion or boson, thus associating the
fold on the string with the degree of freedom corresponding to the particle
in the adjoint representation. Along the same idea, 4D QCD has gluons in the
adjoint representation that can play the role of the folds, as conjectured a
long time ago \cite{ibbardeen}, and used successfully in the phenomenology
of gluon jets in the form of 2D folded strings \cite{anderson}.

\item  The interactions of the folded strings can be inferred from those of
the gauge theory\footnote{%
This can be done by examining the 'tHooft-like QCD equations derived in the
theory of gluons interacting with adjoint fermions. There are two parts in
these equations \cite{ibbardeen} \cite{klebanov}, a zeroth order part that
defines the spectrum, and an interaction part. These can be associated with
operators in the canonical formulation of folds in the lightcone gauge. The
first part describes bound states of $n$ adjoint fermions, and it has a form
that is identical to the {\it quantum eigenvalue equations} for the spectrum
of the operator $P_0^{-}=\sum m_i^2/2p_i^{+}+\gamma |x_i^{-}-x_{i+1}^{-}|$
that describes $n$ folds in string theory in the lightcone gauge, as derived
in \cite{bbhp1}\cite{bbhp2}\cite{barshanson}. The second part, called $%
P_{int}^{-}$, describes interactions among the wavefunctions with different
numbers of adjoint fermions that represent the folds. The string version of
this interaction has not been made precise yet, but is expected to be
similar to the one involving end points \cite{ibequivalence}.}. The old and
new QCD-string interactions are different from those provided by the
Polyakov procedure in standard string theory. Furthermore, there has been a
new attempt at the quantization of folded strings in the path integral
formalism \cite{yankielowicz}. There a prescription is given for including
folds in a generalization of the Polyakov path integral that describes
interacting strings, and a correspondence to QCD\ is explored. This attempt
seems to be related to the approach suggested in \cite{bbhp1}, since the new
modifications of the measure give rise to an effective action that includes
the worldline action for the folds in addition to the string action. The
path integral approach is of interest mainly because it is promising for the
formulation of interacting folded strings. There are many ways of
generalizing the string path integral measure, as well as the action, and
more needs to be done.
\end{itemize}

\noindent These comments summarize our feeling of the past 19 years that, in
addition to being of interest in mathematical physics and generalizations of
string physics, folded strings are important in the further development of
the string-QCD connection.

In this paper we report on further developments in classical 2D string
theory along the direction of \cite{ibjs}. The classical solutions are
useful for interpreting the theory of folded strings in curved space-time
and also helpful for formulating and understanding the quantum theory. In
sections 2,3 we first give new systematic results for the classical folded
string solutions by deriving the general form of the ``transfer operation''
for any 2D target space curved metric, and in sections 4,5,6 we apply the
general formalism explicitly to black hole and cosmological metrics. In the
literature on classical solutions of strings in curved space-time there
exists solutions in higher dimensions which are of a different nature \cite
{devega} than the ones discussed here.

\section{String solutions in 2D curved space-time}

Consider a string $x^\mu (\tau ,\sigma )$ propagating in a 2D curved
space-time manifold characterized by the target space metric $G_{\mu \nu
}(x) $. In the conformal gauge, the classical action is given by $\smallint
d^2\sigma \,G_{\mu \nu }(x)\partial _{+}x^\mu \partial _{-}x^\nu .$ In the
classical theory one can ignore the dilaton (since the dilaton becomes
important for conformal invariance only in the quantum theory at higher
orders of $\hbar $) . In 2D target space-time the antisymmetric tensor $%
B_{\mu \nu }(x)$ can be eliminated since it produces a total derivative in
the action, and the most general metric can always be transformed into the
conformal form $G_{\mu \nu }=\eta _{\mu \nu }G(x).$ Then the most general 2D
{\it classical} string equations of motion and conformal (Virasoro)
constraints can be put into the form
\begin{equation}
\begin{array}{c}
\partial _{+}(G\,\,\partial _{-}u)+\partial _{-}(G\,\,\partial _{+}u)=\frac{%
\partial G}{\partial v}(\partial _{+}u\partial _{-}v+\partial _{+}v\partial
_{-}u) \\
\partial _{+}(G\,\,\partial _{-}v)+\partial _{-}(G\,\,\partial _{+}v)=\frac{%
\partial G}{\partial u}(\partial _{+}u\partial _{-}v+\partial _{+}v\partial
_{-}u) \\
\partial _{+}u\partial _{+}v=0=\partial _{-}u\partial _{-}v\,\,\,,
\end{array}
\label{stringeqs}
\end{equation}
where we have used the target space lightcone coordinates
\begin{equation}
u(\sigma ^{+},\sigma ^{-})=\frac 1{\sqrt{2}}(x^0+x^1),\quad v(\sigma
^{+},\sigma ^{-})=\frac 1{\sqrt{2}}(x^0-x^1),  \label{lightcone}
\end{equation}
and the world sheet lightcone coordinates
\begin{equation}
\sigma ^{\pm }=(\tau \pm \sigma )/\sqrt{2},\,\quad \quad \partial _{\pm
}=(\partial _\tau \pm \partial _\sigma )/\sqrt{2}.  \label{worldlight}
\end{equation}
There is a remaining local conformal invariance of left movers and right
movers under $\sigma ^{\pm }\rightarrow \sigma ^{\prime \pm }(\sigma ^{\pm
}) $ that allows a gauge choice. These equations may be called the ``string
geodesic equations'' since they are indeed the generalization of the
particle geodesic equations
\begin{equation}
\partial _{+}\partial _{-}x^\mu +\Gamma _{\nu \lambda }^\mu \partial
_{+}x^\nu \partial _{-}x^\lambda =0.  \label{geodst}
\end{equation}

Since a typical string state is massive, one should expect that the string
will follow {\it on the average} the trajectory of a {\it massive}
particle.Therefore, to understand the average behavior of the string
geodesic it is useful to first consider the solution for the geodesic of a
massive particle. The particle geodesic equations follow from the above ones
by dimensional reduction. That is, by dropping the $\sigma $ dependence, $%
\partial _{\pm }\rightarrow \partial _\tau ,$ these equations reduce to the
point particle geodesic equations. For particles, the last line in (\ref
{stringeqs}) imposes the condition for a null geodesic, which is too
restrictive for our purpose. If this condition is modified to
\begin{equation}
G\dot u\dot v=\frac{m^2}2
\end{equation}
then (\ref{stringeqs}) become the equations for a timelike geodesic for a
massive particle with mass $m.$ The zero mass limit may also be considered
at the end.

We will provide the explicit solutions to the particle as well as the string
equations. As discovered in \cite{ibjs} there are additional {\it stringy
phenomena due to the wave nature} that cannot be seen in the particle
solution, and therefore it is useful to contrast the string solutions with
the particle solutions.

In seeking classical solutions to the string equations one must impose also
the properties of periodicity and forward propagation that are required on
physical grounds \cite{ibjs}:

\begin{description}
\item  (i) The solution must be periodic in the variable $\sigma ,$ $x^\mu
(\tau ,\sigma )=x^\mu (\tau ,\sigma +4)$,

\item  (ii) Despite the periodicity in $\sigma ,$ the global time coordinate
$T(\tau ,\sigma )=(u+v)/\sqrt{2}$ must always increase as a function of the
proper time $\tau ,$ for all values of $\sigma .$
\end{description}

\noindent  In flat space-time the solutions take the form
\[
x^\mu (\tau ,\sigma )=x_L^\mu (\sigma ^{+})+x_R^\mu (\sigma ^{-}).
\]
One can fix the remaining conformal invariance $\sigma ^{\pm }\rightarrow
\sigma ^{\prime \pm }(\sigma ^{\pm })$ by choosing the gauge, $x_L^0(\sigma
^{+})=q_L^0+p^0\sigma ^{+},\,\,x_R^0(\sigma ^{-})=q_R^0+p^0\sigma ^{-},\,$
such that $x^0\equiv T(\tau ,\sigma )=T_0+p^0\tau .$ This form satisfies
automatically requirement (ii). In curved space-time, in the conformal
gauge, it is not always possible to make $T(\tau ,\sigma )$ only a function
of $\tau ,$ but still, on physical grounds, only the solutions in which it
does not decrease as a function of $\tau $ (for any $\sigma $) can be
admitted. The two properties (i) and (ii) must be maintained for any
physical solution in curved space-time as explained in \cite{ibjs}.

In flat space-time ($G=1$ ) the generalized BBHP\ solution for strings is
\begin{equation}
\begin{array}{c}
u(\sigma ^{+},\sigma ^{-})=u_0+\frac{p^{+}}2\left[ (\sigma ^{+}+f(\sigma
^{+}))+(\sigma ^{-}-g(\sigma ^{-}))\right] \\
v(\sigma ^{+},\sigma ^{-})=v_0+\frac{p^{-}}2\left[ (\sigma ^{+}-f(\sigma
^{+}))+(\sigma ^{-}+g(\sigma ^{-}))\right]
\end{array}
\label{general}
\end{equation}
where $f(\sigma ^{+})$ and $g(\sigma ^{-})$ are any two {\it periodic
functions, } $f(\sigma ^{+})=f(\sigma ^{+}+\sqrt{2})$, $g(\sigma
^{-})=g(\sigma ^{-}+\sqrt{2}),$ with slopes $f^{\prime }(\sigma ^{+})=\pm 1$
and $g^{\prime }(\sigma ^{-})=\pm 1.$ The slope can change discontinuously
any number of times at arbitrary locations $\sigma _i^{+},\sigma _j^{-}$
within the basic intervals $-1/\sqrt{2}\leq \sigma ^{\pm }\leq 1/\sqrt{2}$
(and then repeated periodically), but the functions $f,g$ are continuous at
these points. The discontinuities in the slopes are allowed since the
equations of motion are first order in either $\partial _{+}$ or $\partial
_{-}$. The number of times the slope changes in the basic interval
corresponds to the number of folds for left movers and right movers
respectively. This is seen by taking a snapshot of the string at a constant
value of $T$ (i.e. $\tau =\tau _0=const.$ , in the present case of flat
space-time), which is easily done by plotting the space coordinate of the
string $X(\tau _0,\sigma )=(u-v)/\sqrt{2}$ as a function of $\sigma $. The
plot shows that the string is folded precisely at the points $\sigma _i(\tau
_0)$ where the functions $f,g$ change slopes.

The simplest BBHP solution in flat space-time is the so called yo-yo
solution
\begin{equation}
\begin{array}{c}
u(\sigma ^{+},\sigma ^{-})=u_0+\frac{p^{+}}2\left[ (\sigma ^{+}+\left|
\sigma ^{+}\right| _{per})+(\sigma ^{-}-\left| \sigma ^{-}\right|
_{per})\right] \\
v(\sigma ^{+},\sigma ^{-})=v_0+\frac{p^{-}}2\left[ (\sigma ^{+}-\left|
\sigma ^{+}\right| _{per})+(\sigma ^{-}+\left| \sigma ^{-}\right|
_{per})\right]
\end{array}
\label{old}
\end{equation}
where $f(\sigma ^{+})=|\sigma ^{+}|_{per},\,\,\,g(\sigma ^{-})=-|\sigma
^{-}|_{per}$ are the absolute values of $\sigma ^{\pm }$ taken in the basic
intervals $-1/\sqrt{2}\leq \sigma ^{\pm }\leq 1/\sqrt{2},$ and then repeated
periodically. If $\sigma $ is taken in the full interval $-2\leq \sigma \leq
2$ , then this solution describes the motion of a closed string with two
folds, or if $\sigma $ is taken in the interval $0\leq \sigma \leq 2$ then
it describes an open string without folds but with two end points. The
minimal surface swept by the string for the yo-yo solution is plotted in
Fig.1. For an open string the minimal surface has a single sheet. For a
closed string it consists of two sheets on top of each other, hence folded
at the edges.

As a second example consider the same form as eq.(\ref{old}) but with
different periods for $|\sigma ^{+}|_{per^{+}}$ and $|\sigma ^{-}|_{per^{-}}$%
. For example, take $per^{-}=\frac 1nper^{+}.$ Then there are $n+1$ critical
points that move at the speed of light. In Fig.2 the case of $n=2$, with
three critical points is depicted. Two of these points (the points at the
ends at any $\tau $) are folds, but the third point is a saddle point (in
the plot of $x(\sigma ,\tau _0)$ at fixed $\tau _0$) where the string
attempts to fold. Evidently more folds or critical points are generated by
more complicated choices of $f(\sigma ^{+})$ and $g(\sigma ^{-})$. For
additional plots of more complicated solutions, including a discussion of
relations among different ways of deriving such solutions, the reader should
consult the papers by BBHP.

As suggested in \cite{ibjs}, except for deformations due to curvature and
singularities, the minimal surfaces in curved space-time are analogous, and
they reduce precisely to the BBHP ones in the asymptotically flat regions of
space-time where $G\rightarrow 1.$

In curved space-time the general solution of eqs.(\ref{stringeqs}) fall into
four classes $A,B,C,D$
\begin{equation}
\begin{array}{ll}
A:\qquad u=U(\sigma ^{+}), & \quad v=\bar V(\sigma ^{-}) \\
B:\qquad u=\bar U(\sigma ^{-}), & \quad v=V(\sigma ^{+}) \\
C:\qquad u=u_0, & \quad v=W[\alpha (\sigma ^{+})+\bar \beta (\sigma
^{-}),\,\,u_0] \\
D:\qquad u=\bar W[\bar \alpha (\sigma ^{-})+\beta (\sigma ^{+}),\,\,v_0], &
\quad v=v_0\,\,\,,
\end{array}
\label{fourr}
\end{equation}
where $U(\sigma ^{+}),V(\sigma ^{-})$, $\bar U(\sigma ^{-})$, $\bar V(\sigma
^{+})$, $\alpha (\sigma ^{+})$, $\beta (\sigma ^{+})$, $\bar \alpha (\sigma
^{-})$, $\bar \beta (\sigma ^{-})$ are arbitrary and $u_0,v_0$ are constants%
\footnote{%
This set of solutions were noticed independently in \cite{devega2} and \cite
{ibsfglo} \cite{ibberkeley}, but the authors of \cite{devega2} did not
realize that the validity of these solutions is limited to patches of the
worldsheet, and they assumed that the stringy solutions discussed here and
in \cite{ibjs} are gauged away by using the remaining conformal invariance.}%
. Solutions $A,B$ are present for any metric, but the functions $W,\bar W$
in solutions $C,D$ are obtained by inverting the following relations that
depend on the metric $G$

\begin{equation}
\begin{array}{c}
C:\quad u=u_0,\quad \quad F(u_0,W)\equiv \int^Wdv^{\prime }G(u_0,v^{\prime
})=\alpha (\sigma ^{+})+\bar \beta (\sigma ^{-})\,\,\,, \\
\\
D:\quad v=v_0,\quad \quad \bar F(\bar W,v_0)\equiv \int^{\bar W}du^{\prime
}G(u^{\prime },v_0)=\bar \alpha (\sigma ^{-})+\beta (\sigma ^{+}),
\end{array}
\label{foura}
\end{equation}
where the integration is performed at constant $u=u_0$ for solution $C,$ and
at constant $v=v_0$ for solution $D.$ Taking derivatives $\partial _{\pm }$
of the integrals in eq.(\ref{foura}) gives relations that solve the
equations (\ref{stringeqs}). So, for a given metric $G(u,v)$ there exists
the functions $F(u,v)$ and $\bar F(u,v)$ such that their partial derivatives
reproduce the metric
\begin{equation}
\frac{\partial F(u,v)}{\partial v}=G(u,v)=\frac{\partial \bar F(u,v)}{%
\partial u}.  \label{FGF}
\end{equation}
and for each metric $G$ we have the relations
\begin{equation}
\begin{array}{c}
F(u_0,v)=\alpha +\bar \beta \quad \leftrightarrow \quad v=W(\alpha +\bar
\beta ,u_0), \\
\bar F(u,v_0)=\bar \alpha +\beta \quad \leftrightarrow \quad u=\bar W(\bar
\alpha +\beta ,v_0),
\end{array}
\label{FW}
\end{equation}
that help define the solutions $C,D$ in terms of the arbitrary functions $%
\alpha (\sigma ^{+})$, $\,\beta (\sigma ^{+})$, $\,\bar \alpha (\sigma
^{-}),\,\bar \beta (\sigma ^{-}).$ Consider the following three cases as
illustrations

\begin{enumerate}
\item  Flat metric $ds^2=du\,dv$ :
\begin{equation}
\begin{array}{c}
F=u_0+v=\alpha +\bar \beta ,\quad \leftrightarrow \quad v=W=\alpha +\bar
\beta -u_0, \\
\bar F=u+v_0=\bar \alpha +\beta ,\quad \leftrightarrow \quad u=\bar W=\bar
\alpha +\beta -v_0.
\end{array}
\label{FWflat}
\end{equation}

\item  SL(2,R)/R black hole metric $ds^2=(1-uv)^{-1}du\,dv$ :
\begin{equation}
\begin{array}{l}
F=-u_0^{-1}\ln (1-u_0v)=\alpha +\bar \beta \quad \\
\quad \quad \leftrightarrow \quad v=W=u_0^{-1}\{1-\exp [-u_0(\alpha +\bar \beta
)]\}, \\
\bar F=-v_0^{-1}\ln (1-uv_0)=\bar \alpha +\beta \quad \\
\quad \quad \leftrightarrow \quad u=\bar W=v_0^{-1}\{1-\exp [-v_0(\bar \alpha
+\beta )]\}.
\end{array}
\label{FWsl}
\end{equation}

\item  Cosmological (de Sitter) metric $ds^2=dt^2-e^{2Ht}dx^2=\frac
4{H^2}(u+v)^{-2}du\,dv$:
\begin{equation}
\begin{array}{c}
F=-(u_0+v)^{-1}=\alpha +\bar \beta \quad \leftrightarrow \quad v=W=-(\alpha
+\bar \beta )^{-1}-u_0, \\
\bar F=-(u+v_0)^{-1}=\bar \alpha +\beta \quad \leftrightarrow \quad u=\bar
W=-(\bar \alpha +\beta )^{-1}-v_0.
\end{array}
\label{FWdesitter}
\end{equation}
\end{enumerate}

Since one still needs to impose the periodicity and forward propagation
conditions (i) and (ii) given above, (\ref{fourr}-\ref{FWdesitter}) are not
yet legitimate solutions. As discussed in \cite{ibjs} each one of the forms $%
A,B,C,D$ in (\ref{fourr}) can be valid only in certain patches of the world
sheet ($\sigma ^{+},\sigma ^{-}$), and these solutions need to be matched to
each other at the boundaries of the patches. Thus, to construct a legitimate
solution in curved space-time, first one must decide on the form of the
solution in the flat asymptotic region by making a choice for the functions $%
f,g$ in eq.(\ref{general}). This is a boundary condition which is consistent
with the requirements (i) and (ii). The sign patterns of the derivatives $%
(f^{\prime },g^{\prime })=(+,-),(-,+),(+,+),(-,-)$ divides the world sheet ($%
\sigma ^{+},\sigma ^{-}$) into patches where the corresponding signs hold.
To each such patch the forms $A,B,C,D$ in (\ref{fourr}) are assigned
respectively, for any curved metric $G$. The pattern of assigned forms must
be periodic in the direction of $\sigma $ but not in the direction of $\tau $%
. The patterns $A,B,C,D$ are the same for flat or curved space-time. The
difference between curved and flat space-time arises in the choice of the
metric dependent functions $W,\bar W$ for the $C,D$ patches. Then these
solutions are matched at the boundaries between patches. This procedure
insures the properties (i) and (ii) in curved space-time while being
consistent with boundary conditions (i.e. some given BBHP solution) in the
asymptotically flat region of the target space metric $G$.

It turns out that the constant values of $u$ or $v$ in the $C$ or $D$
patches respectively provide sufficient data for constructing the motion of
the entire string (see below for details). The solutions in these patches
describe the {\it motion of the folds}\footnote{%
By definition, at a fold the determinant of the induced metric, $g_{\alpha
\beta }=\partial _\alpha x^\mu \partial _\beta x^\nu G_{\mu \nu },$
vanishes, $\det g=0$. But in the conformal gauge the induced metric itself
also vanishes locally everywhere since $g_{\alpha \beta }=\Lambda \eta
_{\alpha \beta }$. Note that this does not mean that the world sheet metric $%
\eta _{\alpha \beta }$ vanishes. In the $C,D$ cells by virtue of having
either $u$ or $v$ constant throughout the cell one gets $g_{\alpha \beta
}=0, $ indicating that all points in these cells are mapped to the
trajectory of the fold in target spacetime. The mapping is many to one,
since a region of the world sheet is mapped to a segment (trajectory of the
fold) in target spacetime. Therefore, a fold in target spacetime has many
representatives on the world sheet. For example, consider the leftmost $C$%
-type cell at the bottom of the diagram in (\ref{soll}) for which $%
u=u_{k-1}, $ and $v=W_{k-1}(\sigma ^{+},\sigma ^{-})$. At a constant $\tau
=\tau _0,$ all the $\sigma $ points that give the same value of $v=v_0$ are
mapped to the same fold located at $(u_{k-1},v_0)$. As $\tau $ changes $%
u=u_{k-1}$ remains fixed while $v$ changes along the lightlike trajectory of
the fold. To trace the trajectory of a fold it is sufficient to concentrate
on one of its images on the world sheet. Such representative images are the
vertical lines at $\sigma =0,2$ in the diagram in (\ref{soll}).}.
Specifically note that a constant value of $u$ or $v$ describes a light-like
trajectory, indicating that a fold moves at the speed of light. In order to
determine these constants in different time intervals, one must go through
the procedure of matching boundaries at the patches as defined in the
previous paragraph.

\section{The yo-yo in any curved space-time}

As an illustration of the procedure we consider the simplest boundary
condition in the asymptotic region, namely the yo-yo solution given in (\ref
{old}). The pattern that emerges is as follows. The world{\ \ sheet is
labelled by $\sigma $ horizontally and by $\tau $ vertically. It is sliced
by equally spaced $45^o$ lines that form a light-cone lattice in $\sigma
^{\pm }$. The crosses in the diagram represent the corners of the cells on
the world sheet. Each cell on the world sheet is labelled by the values of $%
(\sigma ^{+},\sigma ^{-})$ at the center of the cell, divided by a factor of
$\sqrt{2}$. For example at the center of the cell labelled by $(m,n)$ the
world sheet coordinates are $\sigma ^{+}=m\sqrt{2},\ \ \ \sigma ^{-}=n\sqrt{2%
}$, and the $(\tau ,\sigma )$ coordinates are $\tau =m+n,\quad \sigma =m-n$.
The points inside the cell $(m,n)$ are parametrized by $\sigma ^{\pm }$ in
the ranges
\begin{equation}
(m-{\frac 12})\sqrt{2}<\sigma ^{+}<(m+{\frac 12})\sqrt{2},\qquad (n-{\frac 12%
})\sqrt{2}<\sigma ^{-}<(n+{\frac 12})\sqrt{2}.  \label{means}
\end{equation}
The }$A,B,C,D$ solutions that are placed into these cells provide a map from
the world sheet to the target space-time.

{\tiny
\begin{equation}
\begin{array}{ccccc}
\begin{array}{c}
\sigma =0 \\
\vdots
\end{array}
&
\begin{array}{c}
\sigma =1 \\
\vdots
\end{array}
&
\begin{array}{c}
\sigma =2 \\
\vdots
\end{array}
&
\begin{array}{c}
\sigma =3 \\
\vdots
\end{array}
&
\begin{array}{c}
\sigma =4\equiv 0 \\
\vdots
\end{array}
\\
\times &
\begin{array}{c}
\begin{array}{c}
U_{k+2}(\sigma ^{+}) \\
V_{k+2}(\sigma ^{-})
\end{array}
\end{array}
& \times &
\begin{array}{c}
U_{k+2}(\sigma ^{-}) \\
V_{k+2}(\sigma ^{+})
\end{array}
& \times \\
&  &  &  &  \\
\cdots
\begin{array}{c}
u_{k+1} \\
W_{k+1}(\sigma ^{+},\sigma ^{-})
\end{array}
& \times &
\begin{array}{c}
\bar W_{k+1}(\sigma ^{+},\sigma ^{-}) \\
v_{k+1}
\end{array}
& \times &
\begin{array}{c}
u_{k+1} \\
W_{k+1}(\sigma ^{+},\sigma ^{-})
\end{array}
\cdots \\
&  &  &  &  \\
\times &
\begin{array}{c}
U_{k+1}(\sigma ^{-}) \\
V_{k+1}(\sigma ^{+})
\end{array}
& \times &
\begin{array}{c}
U_{k+1}(\sigma ^{+}) \\
V_{k+1}(\sigma ^{-})
\end{array}
& \times \\
&  &  &  &  \\
\cdots
\begin{array}{c}
\bar W_k(\sigma ^{+},\sigma ^{-}) \\
v_k
\end{array}
& \times &
\begin{array}{c}
u_k \\
W_k(\sigma ^{+},\sigma ^{-})
\end{array}
& \times &
\begin{array}{c}
\bar W_k(\sigma ^{+},\sigma ^{-}) \\
v_k
\end{array}
\cdots \\
&  &  &  &  \\
\times &
\begin{array}{c}
U_k(\sigma ^{+}) \\
V_k(\sigma ^{-})
\end{array}
& \times &
\begin{array}{c}
U_k(\sigma ^{-}) \\
V_k(\sigma ^{+})
\end{array}
& \times \\
&  &  &  &  \\
\cdots
\begin{array}{c}
u_{k-1} \\
W_{k-1}(\sigma ^{+},\sigma ^{-})
\end{array}
& \times &
\begin{array}{c}
\bar W_{k-1}(\sigma ^{+},\sigma ^{-}) \\
v_{k-1}
\end{array}
& \times &
\begin{array}{c}
u_{k-1} \\
W_{k-1}(\sigma ^{+},\sigma ^{-})
\end{array}
\cdots \\
\quad \vdots & \vdots & \quad \quad \vdots \quad \quad & \vdots & \quad
\vdots \quad
\end{array}
\label{soll}
\end{equation}
} {\ Consider all the cells in a horizontal row corresponding to a fixed
value of $\tau .$ There are two types of cells, the $A,B$ type whose centers
are at $\tau =m+n=2k$ (i.e. $m=k+l,\,\,n=k-l$)$,$ and the $C,D$ type cells
whose centers are at $\tau =m+n=2k+1$ (i.e. $m=k+l+1,\,\,n=k-l\,\,$ or $%
\,\,m=k+l,\,n=k-l+1$). Corresponding to the $A,B,C,D$ patterns, the
solutions in (\ref{fourr}) are assigned periodically in $\sigma $ as
follows: }

\begin{itemize}
\item  {\ {All the $A$ cells whose centers are at $\tau =2k$ , $m=even,$ $%
n=even$ are assigned periodically the same solution $u=U_k(\sigma
^{+}),\,v=V_k(\sigma ^{-})$,} }

\item  {\ {All the $B$ cells whose centers are at $\tau =2k$ , $m=odd,$ $%
n=odd$ are assigned periodically the same solution\footnote{{\ {\ A priori
the $U,V$ solutions are represented by different functions in the $A,B$
patches at the same $\tau $. The convenient use of the same set of functions
$U_k(z),V_k(z)$ for both the $A,B$ type cells (but with $z=\sigma ^{\pm
}\rightarrow \sigma ^{\mp })$ corresponds to fixing the remaining conformal
gauge partially. One is allowed to choose a gauge locally as long as this is
not in conflict with the matching of boundary conditions.}}} $u=U_k(\sigma
^{-}),\,v=V_k(\sigma ^{+}),$ } }

\item  {\ {\ All the $C$ cells whose centers are at $\tau =2k+1$ , $m=odd,$ $%
n=even$ are assigned periodically the same solution $u=u_k\,,\,v=W_k(\sigma
^{+},\sigma ^{-}),$ } }

\item  {\ {\ All the $D$ cells whose centers are at $\tau =2k+1$ , $m=even,$
$n=odd$ are assigned periodically the same solution $u=\bar W_k(\sigma
^{+},\sigma ^{-}),\,v=v_k$ } }
\end{itemize}

{\ {\ \noindent By assigning the {\it same function} to all the cells of the
same type at a fixed $\tau $ (or fixed $k$) one obtains a pattern that
insures periodicity under $\sigma \rightarrow \sigma +4,$ (or $l\rightarrow
l+2).$ This periodicity may also be insured by taking periodic functions $%
U_k(z+\sqrt{2})=U_k(z)$ and $V_k(z+\sqrt{2})=V_k(z)$. For different $k$
(i.e. different $\tau $) the functions $U_k,V_k,$ etc. are different, but
are related to each other by matching boundary conditions across the cell
boundaries. Therefore, this procedure corresponds to a world-sheet with the
topology of a cylinder. The map provided by the functions is from the
cylinder to curved space-time whose metric is $G_{\mu \nu }=\eta _{\mu \nu
}G(u,v).$ } }

{\ {\ The continuity at the corners that join the $A,B$ cells is
automatically insured by the use of the same functions $U_k(z),V_k(z)$ to
describe the $A,B$ solutions (but with different arguments $z=\sigma ^{\pm }$
that alternate between neighboring cells, see footnote). Continuity at the
boundaries between $A,B$ cells and $C,D$ cells requires
\begin{equation}
\begin{array}{c}
U_{k+1}(-1/\sqrt{2})=U_k(1/\sqrt{2})=u_k, \\
V_{k+1}(-1/\sqrt{2})=V_k(1/\sqrt{2})=v_k.
\end{array}
\label{boundary}
\end{equation}
where the $(u_k,v_k)$ are constants. Similarly, by taking into account the
relations (\ref{FW}) at these boundaries one can construct the functions $%
W_k,\bar W_k$ for the $C,D$ cells in terms of the functions $U_k,V_k$%
\begin{equation}
\begin{array}{l}
W_k(\sigma ^{+},\sigma ^{-})=W\left[ \left( F(u_k,V_k(\sigma
^{+}))+F(u_k,V_k(\sigma ^{-}))-F(u_k,v_{k-1})\right) \,,u_k\right] \\
\bar W_k(\sigma ^{+},\sigma ^{-})=\bar W\left[ \left( \bar F(U_k(\sigma
^{+}),v_k)+\bar F(U_k(\sigma ^{-}),v_k)-\bar F(u_{k-1},v_k)\right)
\,,v_k\right] .
\end{array}
\label{wwbar}
\end{equation}
Evaluating these at the lower (i.e. past) boundaries of the $C,D$ cells,
using $V_k(-1/\sqrt{2})=v_{k-1},$ $U_k(-1/\sqrt{2})=u_{k-1}$ , the boundary
matching is insured by the fact that $F$ and $W$ are inverses of each other
(see eq.(\ref{FW}-\ref{FWdesitter})
\begin{equation}
W\left( F(u_k,V_k(z)),u_k\right) =V_k(z),  \label{lower}
\end{equation}
and similarly for $U_k(z)$. At the upper (i.e. future) boundaries of the $%
C,D $ cells the boundary matching gives a {\it recursion relation}
\begin{equation}
\begin{array}{l}
V_{k+1}(z)=W\left[ \left( F(u_k,V_k(z))+F(u_k,v_k)-F(u_k,v_{k-1})\right)
\,,u_k\right] \\
U_{k+1}(z)=\bar W\left[ \left( \bar F(U_k(z),v_k)+\bar F(u_k,v_k)-\bar
F(u_{k-1},v_k)\right) \,,v_k\right] ,
\end{array}
\label{recursion}
\end{equation}
where $z=\sigma ^{\pm }.\,$This recursion may be viewed as a {\it transfer}
\thinspace operation in proper time $\tau \rightarrow \tau +2$, for any $%
\sigma $, and is quite analogous to the concept of the ``transfer matrix''
in lattice theories. The recursion leads to the solution of all the $%
U_k(\sigma ^{\pm }),\,V_k(\sigma ^{\pm })$ in terms of $U_0(z),\,V_0(z)$,
that describe initial conditions at $\tau =0$. } }

{\ {\ By evaluating the recursion relation (\ref{recursion}) at the
boundaries of each cell $z=\pm 1/\sqrt{2}$ and using the values (\ref
{boundary}) at the boundaries, one finds a recursion relation for the
constants $(u_k,v_k)$%
\begin{equation}
\begin{array}{l}
v_{k+1}=W\left[ \left( 2F(u_k,v_k)-F(u_k,v_{k-1})\right) \,,u_k\right] , \\
u_{k+1}=\bar W\left[ \left( 2\bar F(u_k,v_k)-\bar F(u_{k-1},v_k)\right)
\,,v_k\right] .
\end{array}
\label{constants}
\end{equation}
The solution of this recursion relation requires $4$ initial constants $%
u_0,v_0$, $u_{-1}$, $v_{-1}$
\begin{equation}
\begin{array}{c}
U_0(-1/\sqrt{2})=u_{-1}\quad U_0(1/\sqrt{2})=u_0\quad , \\
V_0(-1/\sqrt{2})=v_{-1}\quad V_0(1/\sqrt{2})=v_0\quad .
\end{array}
\label{init}
\end{equation}
Therefore, the positions $(u_k,v_k)$ are fully determined in curved
space-time in terms of 4 initial constants. } }

{\ {\ The constants $(u_k,v_k)$ are sufficient to describe the physical
motion of the folds (or end points), as well as the whole string, as
follows. Consider the diagram of eq.(\ref{soll}). At any $\tau $ the
trajectories of the folds are parametrized by the vertical lines that pass
through $\sigma =0,2$ on the world sheet (and their periodic repetitions at\
$\sigma =4l,\,4l+2,$ see footnote$)$. Likewise, vertical lines that pass
through the crosses located at $\sigma =1,3$ (and their periodic repetitions
at $\sigma =4l+1,\,4l+3)$ parametrize the trajectory of the midpoint between
the folds. The center of mass of the string coincides with these midpoints.
As $\tau $ increases one can read off the space-time trajectories of the
center of mass and of the folds by moving upward along the vertical lines in
the diagram. For example, consider the $\sigma =0$ fold: during 2$k-2\leq
\tau \leq 2k$ it remains at constant $u=u_{k-1}$ while the value of $%
v=W_{k-1}$ increases from $v=v_{k-2}$ to $v=v_k$. Between 2$k\leq \tau \leq
2k+2$ it remains at constant $v=v_k$ while the value of $u=\bar W_k$
increases from $u=u_{k-1}$ to $u=u_{k+1},$ etc. In a similar way the
trajectory of the second fold and of the center of mass are read off
directly from the diagram in eq.(\ref{soll}). The space-time trajectories of
these points are plotted in a $(u,v)$ plot in Fig.3. } }

{\ {\ The detailed motion of the intermediate points of the string at any $%
\sigma $ are described by the functions $U_k,V_k,W_k,\bar W_k$ as indicated
on the diagram (\ref{soll}) and mapped on Fig.3. The space-time trajectories
of folds or end points that are the images of $\sigma =0,2$ are physical and
cannot depend on conformal reparametrizations. Indeed, as seen from the
above solution there is no freedom in the choice of the constants $(u_k,v_k)$
except for the initial values (\ref{init}). On the other hand, the motion of
the rest of the string is gauge dependent at intermediate points $\sigma $
(because of reparametrizations), and therefore it depends on the choice of $%
U_0(z),V_0(z)$ that have remained unspecified. However, once the motion of
the end points is plotted, it is clear from Fig.3 that the shape of the
minimal surface is already determined without needing the details of the
gauge dependent motion of the intermediate points. } }

{\ {\ The remaining conformal invariance may be used to fix the form of
these functions in the initial cell (although this is not necessary). For
the yo-yo solution the initial functions $U_0(z),\,V_0(z)$ need not contain
more than $4$ constants that are related to the initial positions and
velocities of the two folds. Therefore, the simplest gauge fixed form is
\begin{equation}  \label{initial}
\begin{array}{c}
U_0(z)=\frac 12(u_0+u_{-1})+\frac 1{\sqrt{2}}(u_0-u_{-1})\,z_{per} \\
V_0(z)=\frac 12(v_0+v_{-1})+\frac 1{\sqrt{2}}(v_0-v_{-1})\,z_{per},
\end{array}
\end{equation}
where $z_{per}$ is the linear function $z_{per}=z$ in the interval $-1/\sqrt{%
2}\leq z\leq 1/\sqrt{2}$, and then repeated periodically. However, any other
periodic function with the same 4 boundary constants will produce the same
physical motion for the folds. } }

{\ {\ The recursion (\ref{constants}) is the fundamental physical relation
that fully determines the motion of the yo-yo string in curved space-time.
We called it the ``transfer matrix'' in the example of the black hole worked
out in ref.\cite{ibjs}. It was found that it has certain invariances that
are valid everywhere in target space-time, including near singularities. The
invariance is related to a lattice version of the fundamental action $%
A=\smallint d^2\sigma \,G_{\mu \nu }\partial _{+}x^\mu \partial _{-}x^\nu $
that represents the minimal surface swept by the string. The lattice version
of the minimal surface is expressed in terms of the constants $(u_k,v_k)$ ,
and its value for one period turns out to be a constant of motion. Explicit
expressions for this lattice action will be given for specific metrics in
the following sections. For every metric $G$ one can find a lattice version
of the action $A$ that is an invariant under the recursion (\ref{constants}%
). The invariance is valid even in the vicinity of singularities in
space-time ( i.e. when $G(u_k,v_k)$ grows) and helps in the understanding of
new stringy phenomena. For example, it was found that classical strings can
tunnel to regions of space-time (such as the bare singularity region of a
black hole) that are forbidden to particle geodesics. Such a surprising
motion of a string may be thought of as the analog of the diffraction of
light around corners, that is possible for classical waves, but is
impossible for particle trajectories. } }

{\ {\ In this section we constructed the yo-yo solution in any curved
space-time given by $G$. In a similar way one may consider more complicated
solutions with many folds. The general boundary condition near $G\rightarrow
1$ given by (\ref{general}), with any number folds, defines a pattern of $%
A,B,C,D$ on the world sheet that corresponds to the regions of $(\sigma
^{+},\sigma ^{-})$ that have definite signs of $f^{\prime },g^{\prime }$ for
some choice of $f,g.$ The pattern must be periodic horizontally, with a
period of $\sigma \rightarrow \sigma +4$, to insure periodicity. This
generalizes the lattice in the diagram of (\ref{soll}). By virtue of the
BBHP construction, any of these generalized patterns is guaranteed to
correspond to strings that propagate forward in time. Then there remains to
carry out the matching of the functions at the boundaries. This would give
generalizations of the recursion relations and transfer matrices discussed
above. It seems that this is a very rich area for mathematical physics,
since one may explore relations between geometries defined by metrics $G$,
lattices, and transfer matrices. It is clear that the general behavior of
the minimal surface that emerges from this procedure has to be quite similar
to the one in flat space-time (which is already given by the choice of $f,g$%
), except for the deformations due to curvature and singularities. Moreover,
it seems that the main physical stringy features related to the curvature
and/or singularity structure of space-time may already be extracted from the
yo-yo solution that has only two folds. } }

{\ {\ We now apply the general yo-yo results to several specific metrics and
construct explicitly the corresponding ``transfer matrices'', their
invariants, and the corresponding string solutions. } }

\section{Flat Space-time}

{\ {\ The functions $\bar F,\bar W$ corresponding to the flat space-time
metric $G=1$ are given in (\ref{FWflat}). Using them in the general formulas
(\ref{boundary}-\ref{init}) we obtain the explicit recursion relations
\begin{equation}
\begin{array}{c}
\bar W_k=U_k(\sigma ^{+})+U_k(\sigma ^{-})-u_{k-1}, \\
U_{k+1}(z)=U_k(z)+u_k-u_{k-1} \\
u_{k+1}=2u_k-u_{k-1}
\end{array}
\label{recflat}
\end{equation}
They are solved by
\begin{equation}
\begin{array}{c}
u_k=u_0+k(u_0-u_{-1}) \\
U_k(z)=U_0(z)+k(u_0-u_{-1}) \\
\bar W_k=U_0(\sigma ^{+})+U_0(\sigma ^{-})+(k+1)(u_0-u_{-1})-u_0
\end{array}
\label{solflat}
\end{equation}
where $U_0(-1/\sqrt{2})=u_{-1},\,\,U_0(1/\sqrt{2})=u_0$, and the function $%
U_0(z)$ is arbitrary. The solutions for $V_k(z),W_k,v_k$ are obtained from
the above by replacing $U\rightarrow V$ and $u\rightarrow v$. If $%
U_0(z),V_0(z)$ are gauge fixed as in (\ref{initial}), then this solution
takes the convenient form of the BBHP yo-yo string in (\ref{old}). The
present form is a generalization that permits other gauge choices. The
motion of the end points, as plotted in Fig.1 is gauge independent, but the
motion of the interior points of the string depends on the gauge choice, as
expected. } }

{\ {\ Define a lattice version of the surface element $dA=d^2\sigma
\,(\partial _{+}u\,\partial _{-}v+\partial _{-}u\,\partial _{+}v)$ swept by
the string during $2k\leq \tau \leq 2k+2.$ The area of one rectangle in
Fig.1 is
\begin{equation}
dA_k=(u_k-u_{k-1})\,(v_k-v_{k-1}).  \label{minimflat}
\end{equation}
{}From the world sheet point of view this covers the image of one $A$ or $B$
cell, while the image of a $C$ or $D$ cell has zero area in target
space-time (since they are mapped to the edges of the rectangle). Consider
the transformation (\ref{recflat}) as a transfer matrix that takes the
system forward in time. Under this transformation $dA_k$ is an invariant
since $dA_{k+1}=dA_k$ . This is seen by rewriting (\ref{recflat}) in the
form $U_{k+1}(z)-u_k=U_k(z)-u_{k-1}$, etc.. Therefore, we may say that the
``transfer matrix'' for flat space-time given by (\ref{recflat}) leaves
invariant the ``lattice action density'' given by (\ref{minimflat}). This
concept generalizes to curved space-time, as seen below. } }

\section{Black hole space-time}

{\ {\ The case of the SL(2,R)/R two dimensional black hole metric $%
ds^2=(1-uv)^{-1}du\,dv$ was already discussed in \cite{ibjs}, but here we
will show how the results of \cite{ibjs} follow from the general formulas,
and also give the additional recursion relations for $U_k,V_k,W_k,\bar W_k$
at general $k$ and general gauge that were not provided in \cite{ibjs}. } }

{\ {\ The solution for the geodesic of a massive particle was given in our
previous work \cite{ibsfglo} \cite{ibjs}. Here we rewrite it in a more
convenient form
\begin{equation}
\begin{array}{c}
u(\tau )=e^{\sqrt{\gamma ^2+m^2/2}\,\tau }\left[ u_0\cosh (\gamma \tau
)-\left( u_0\sqrt{\gamma ^2+m^2/2}-\dot u_0\right) \frac 1\gamma \sinh
(\gamma \tau )\right] \\
v(\tau )=e^{-\sqrt{\gamma ^2+m^2/2}\,\tau }\left[ v_0\cosh (\gamma \tau
)+\left( v_0\sqrt{\gamma ^2+m^2/2}+\dot v_0\right) \frac 1\gamma \sinh
(\gamma \tau )\right] ,
\end{array}
\label{geosl}
\end{equation}
where $u_0,v_0,\dot u_0,\dot v_0$ are initial velocities and momenta, $m$ is
the mass of the particle, and $\gamma $ is a convenient parameter
\begin{equation}
\gamma =\frac{\sqrt{(u_0\dot v_0+\dot u_0v_0)^2-4\dot u_0\dot v_0}}{%
2(1-u_0v_0)},\quad \frac{\dot u_0\dot v_0}{(1-u_0v_0)}=\frac{m^2}2.
\label{geoslpar}
\end{equation}
In the zero mass limit either $\dot u_0=0$ or $\dot v_0=0$, and then the
solution reduces to a light-like geodesic for which either $u$ or $v$ remain
constant respectively at all times. } }

{\ {\ The singularity is at $u(\tau )v(\tau )-1=0.$ To see when the particle
hits the singularity we compute this quantity
\begin{equation}
\frac{u(\tau )v(\tau )-1}{u_0v_0-1}=\left[ \cosh \gamma \tau +\frac{u_0\dot
v_0+\dot u_0v_0}{\sqrt{(u_0\dot v_0+\dot u_0v_0)^2-4\dot u_0\dot v_0}}\sinh
\gamma \tau \right] ^2.  \label{hit}
\end{equation}
In the massless limit this expression becomes
\begin{equation}
\frac{u(\tau )v(\tau )-1}{u_0v_0-1}=\exp \left( (u_0\dot v_0+\dot
u_0v_0)\tau \right) ,\quad \dot u_0\dot v_0=0.  \label{hitt}
\end{equation}
It is evident that the sign of $uv-1$ cannot change as $\tau $ changes,
therefore the particle must remain in either the black hole region $uv<1$
(can cross the horizon at $u=0$ or $v=0$), or in the bare singularity region
$uv>1.$ The boundary $uv=1$ acts like an impenetrable wall from either side.
This last feature is different for the string solution. In contrast to the
point particle, the string will tunnel through the wall !! This surprising
effect was discovered in \cite{ibjs}. } }

{\ {\ It was evident from the work of \cite{ibjs} that, except for the
tunneling type phenomena, the string follows more or less the geodesic of
the {\it massive} particle. Therefore, it is useful to clarify the
properties of the geodesics of the point particle, because they depend on
the initial particle location as well as its velocity. } }

\begin{itemize}
\item  {\ {\ If the particle starts out in the ``bare singularity'' region ,
$u_0v_0>1$ (future or past regions)$,$ the mass formula in (\ref{geoslpar})
requires $\dot u_0\dot v_0<0$ and $\gamma $ is real$.$ Then the motion is
governed by hyperbolic functions, and (\ref{hit}) never vanishes. Therefore,
a {\it massive} {\it particle, or the string, cannot hit the singularity}.
In the massless limit, according to (\ref{hitt}), the light-like geodesic
will hit the bare singularity only if it starts out with initial conditions
that give $u_0\dot v_0+\dot u_0v_0<0,$ but in any case it reaches the
singularity only at infinite proper time $\tau =\infty .$ Therefore, the
``bare singularity'' region of the SL(2,R)/R black hole is not a singularity
that can be reached by physical signals in a finite amount of proper time.
In this sense it is {\it not really a singularity}. } }

\item  {\ {\ If the particle starts initially in the black hole region $%
u_0v_0<1,\,$ either inside or outside the horizon$,$ its trajectory has
wildly different behavior depending on its velocity. There are two critical
ratios of the velocities at which $\gamma =0.$ } }

\begin{description}
\item  {\ {\ (i) If the velocities lie in the range
\begin{equation}
\left( \frac{1-\sqrt{1-u_0v_0}}{u_0}\right) ^2<\frac{\dot v_0}{\dot u_0}%
<\left( \frac{1+\sqrt{1-u_0v_0}}{u_0}\right) ^2.  \label{critical}
\end{equation}
then $\gamma $ is imaginary and (\ref{hit}) vanishes periodically. The
massive particle goes through the horizon and hits the future singularity at
a {\it finite} value of $\tau .$ There it moves smoothly to a second sheet
of the $(u,v)$ space-time, but still with $uv<1$. It continues its journey
toward the second branch of the singularity and hits it, moving on to a
third sheet of space-time (or back to the first sheet, according to
interpretation). The journey continues endlessly from singularity to
singularity, always moving smoothly to another sheet, and always remaining
in the region $uv<1$. This behavior is similar to the behavior of geodesics
in the many worlds of the Reissner-Nordtrom black hole\footnote{{\ {\ In the
present case the worlds are pasted to each other just at $uv=1$ along the
singularity. When the metric is modified by quantum corrections \cite{ibsf}
a gap develops so that the singularity becomes unreachable while the
geodesics move from one world to the next.}}}. } }

{\ {\ (ii) If the velocities lie in the range
\begin{equation}
\frac{\dot v_0}{\dot u_0}>\left( \frac{1+\sqrt{1-u_0v_0}}{u_0}\right) ^2
\label{criticalii}
\end{equation}
then $\gamma $ is real, the motion is hyperbolic, and (\ref{hit}) vanishes
only once. Therefore, the particle hits the black hole at a finite $\tau $
only once, and moves to a second sheet where it remains for the rest of
time. } }

\item  {\ {\ (iii) If the velocities lie in the range
\begin{equation}
\frac{\dot v_0}{\dot u_0}<\left( \frac{1-\sqrt{1-u_0v_0}}{u_0}\right) ^2
\label{criticaliii}
\end{equation}
then $\gamma $ is real, the motion is hyperbolic, but (\ref{hit}) never
vanishes. Therefore, the particle never hits the black hole. } }
\end{description}
\end{itemize}

{\ {\ The string geodesics given below follows, on the average, the behavior
of the massive particle geodesics above. But, because of the oscillatory
motion we find new phenomena in the vicinity of the black hole. When the
string approaches the black hole from the $uv<1$ region, and hits the
singularity, it behaves differently than the particle: it fully penetrates
the wall to the $uv>1$ region, but then it snaps back into the $uv<1$
region, and then follows more or less the particle trajectory in the second
sheet, etc. (see the solution below and the plots in Figs.4,5). } }

{\ {\ To construct the string solution we use the general formulas of the
previous sections. The functions $\bar F,\bar W$ corresponding to the flat
space-time metric $G=(1-uv)^{-1}$ are given in (\ref{FWflat}). Using them in
the general formulas (\ref{boundary}-\ref{init}) we obtain the explicit
recursion relations
\begin{equation}
\begin{array}{c}
\bar W_k=\frac 1{v_k}\left[ 1-\frac{\left( 1-U_k(\sigma ^{+})v_k\right) \
\left( 1-U_k(\sigma ^{-})v_k\right) }{1-u_{k-1}v_k}\right] \\
U_{k+1}(z)=\frac{1-u_kv_k}{1-u_{k-1}v_k}\left[ U_k(z)+\frac{u_k-u_{k-1}}{%
1-u_kv_k}\right] \\
u_{k+1}=\frac{2u_k-u_{k-1}-u_k^2v_k}{1-u_{k-1}v_k},
\end{array}
\label{recsl}
\end{equation}
and similarly $W_k,V_k,v_k$ are obtained from the above by interchanging $%
U\leftrightarrow V$ and $u\leftrightarrow v.$ This agrees with the results
of \cite{ibjs}. Note that for $u,v\rightarrow 0$ or $\infty $ the metric
approaches the flat metric. In both of these limits the formulas in (\ref
{recsl}) approach the flat ones in (\ref{recflat}). } }

{\ {\ By feeding the recursion relations to a computer, the trajectories of
the folds are plotted in Fig.4,5. A physical discussion of the string
falling into a black hole was given in \cite{ibjs}. The most surprising
effect was the tunnelling of the string into the bare singularity region
which is not possible for particles (Fig.5). As suggested before, this is
analogous to the diffraction of classical light waves that is possible for
waves but not for particles. } }

{\ {\ Just as the flat case, we define a lattice version of the area element
in curved space-time. The ``lattice area'' swept by the string for one of
the rectangles in Fig.4,5 is defined as
\begin{equation}
dA_k=\frac{(u_k-u_{k-1})\,(v_k-v_{k-1})}{1-\frac
14(u_k+u_{k-1})\,(v_k+v_{k-1})}.  \label{minimsl}
\end{equation}
As in the flat case, this is a lattice version of the target space area of
the image of a $A$ or $B$ cell on the world sheet, while the area of the
image of a $C$ or $D$ cell is zero. This expression is invariant under the
``transfer matrix'' (\ref{recsl}), i.e. $dA_k=dA_{k+1}$. The invariance of
this expression everywhere, including in the vicinity of the black hole
singularity, is helpful in understanding the reason for the tunnelling to
the bare singularity region. Namely, since the string must move in a way
that conserves this minimal area, and must have a continuous trajectory, it
cannot avoid the tunnelling for generic initial conditions set by an
observer (see Fig.5). } }

\section{Cosmological space-time}

{\ {\ Consider the cosmological space-time corresponding to a Friedman -
Robertson - Walker (FRW) universe in 4D
\begin{equation}
ds^2=dt^2-R^2(t)\,\left( \frac{dr^2}{1-kr^2}+r^2d\Omega ^2\right) .
\label{FRW}
\end{equation}
where $k=-1,0,1$ are related to the classification of cosmological
space-times as ``open, flat, closed'' respectively. For a string moving
purely along the radial direction $d\theta =d\phi =0$ one concentrates on
the 2D metric
\begin{equation}
ds^2=dt^2-R^2(t)\frac{dr^2}{1-kr^2}.  \label{frw2d}
\end{equation}
It is convenient to change variables
\begin{equation}
\begin{array}{c}
\sqrt{k}\,r=\sin (\sqrt{k}X),\quad T=\int^t\frac{dt^{\prime }}{R(t^{\prime })%
},\quad \sqrt{k}=i,0,1 \\
u=\frac 1{\sqrt{2}}(T+X),\quad v=\frac 1{\sqrt{2}}(T-X),
\end{array}
\label{rX}
\end{equation}
so that the line element takes the conformal form
\begin{equation}
\begin{array}{c}
ds^2=R^2\left( dT^2-dX^2\right) =G(u,v)\,dudv, \\
G(u,v)=2R^2(t(T)).
\end{array}
\label{cosmoconf}
\end{equation}
Once written in terms of $(u,v)$ the complete manifold is usually obtained
by analytic continuation to all values of these variables. Then one may
apply the general formulas of the previous sections to obtain the classical
motion of strings. } }

{\ {\ As an example consider the de Sitter universe for which the expansion
factor of the universe is given by
\begin{equation}  \label{expan}
|R(t)|=e^{Ht}
\end{equation}
where $H=\dot R/R$ is the Hubble constant, and
\begin{equation}  \label{desit}
ds^2=\frac{4\,du\,dv}{H^2(u+v)^2}.
\end{equation}
This 2D space can be embedded in 3D as the surface of a hyperboloid
described by
\begin{equation}  \label{D23D}
\begin{array}{c}
x_0^2-x_1^2-x_2^2=-H^{-2} \\
x_0=\frac{uv-H^{-2}}{u+v},\quad x_1=\frac{uv+H^{-2}}{u+v},\quad x_2=\frac 1H%
\frac{u-v}{u+v}
\end{array}
\end{equation}
Then the metric in (\ref{desit}) takes the flat form
\begin{equation}  \label{flatdesit}
ds^2=dx_0^2-dx_1^2-dx_2^2.
\end{equation}
} }

{\ {\ First consider the geodesic equations for a massive particle of mass $%
m $ . They can be solved exactly as a function of proper time $\tau $%
\begin{equation}  \label{soldesit}
\begin{array}{c}
u(\tau )=c+ \frac{\sinh (Hm\tau )-\sinh (Hm\tau _0)}{H\sinh [Hm(\tau +\tau
_0)]} \\
v(\tau )=-c- \frac{\sinh (Hm\tau )+\sinh (Hm\tau _0)}{H\sinh [Hm(\tau +\tau
_0)]} \\
R(\tau )=\frac{\sinh [Hm(\tau +\tau _0)]}{\sinh (Hm\tau _0)}=-\frac 2H\frac
1{u+v}
\end{array}
\end{equation}
where $c,\tau _0$ are constants, and $R(\tau =0)=1$ has been chosen for
simplicity. The geodesic for the massive particle is best pictured on the
surface of the hyperboloid $x_1^2+x_2^2=x_0^2+H^{-2}$. Inserting the
solution in (\ref{D23D}) one sees that $x_0(\tau )$ increases monotonically
and lies in the range $-\infty <x_0(\tau )<\infty $ . The geodesic extends
from a point on the infinitely large circle at $x_0=-\infty $ to a point on
the infinitely large circle at $x_0=\infty .$ It is a line that spirals less
than or equal to one time on this surface. Define the angle $\tan \theta
=x_2/x_1.$ If the mass is zero, the maximum spiralling angle $\Delta \theta
=\theta (\infty )-\theta (-\infty )$ is exactly $2\pi ,$ but for the massive
particle the angle is less than $2\pi .$ } }

{\ {\ Thus, on the average, we must expect the string center of mass to
spiral less than $2\pi .$ Of course, the overall string performs the yo-yo
oscillations of Fig.3 and sweeps a minimal surface on the hyperboloid, that
is similar to the one in flat space-time except for deformations due to
curvature. } }

{\ {\ The explicit solution that describes this motion is obtained by
applying our general procedure that yields the transfer matrix
\begin{equation}
\begin{array}{c}
\bar W_k(\sigma ^{+},\sigma ^{-})=\left[ \frac 1{U_k(\sigma ^{+})+v_k}+\frac
1{U_k(\sigma ^{-})+v_k}-\frac 1{u_{k-1}+v_k}\right] ^{-1}-v_k \\
U_{k+1}(z)=\left[ \frac 1{U_k(z)+v_k}+\frac 1{u_k+v_k}-\frac
1{u_{k-1}+v_k}\right] ^{-1}-v_k \\
u_{k+1}=\left[ \frac 2{u_k+v_k}-\frac 1{u_k+v_{k-1}}\right] ^{-1}-v_k
\end{array}
\label{transdesit}
\end{equation}
Similar formulas hold for $W_k,V_k,v_k$ respectively. We define a discrete
version of the minimal area for rectangle $k$ by
\begin{equation}
dA_k=\frac 4{H^2}\frac{(u_k-u_{k-1})\,(v_k-v_{k-1})}{%
(u_k+v_{k-1})(u_{k-1}+v_k)}.  \label{minimdesit}
\end{equation}
The transfer matrix leaves invariant this discrete minimal area, i.e. $%
dA_{k+1}=dA_k$. This is easily proven by rewriting the transfer matrix in
the form
\begin{equation}
\begin{array}{c}
\frac{U_{k+1}(z)-u_k}{[U_{k+1}(z)+v_k](u_k+v_k)}=\frac{U_k(z)-u_{k-1}}{%
[U_k(z)+v_k](u_{k-1}+v_k)} \\
\\
\frac{V_{k+1}(z)-v_k}{[u_k+V_{k+1}(z)](u_k+v_k)}=\frac{V_k(z)-v_{k-1}}{%
[u_k+V_k(z)](u_k+v_{k-1})}.
\end{array}
\label{transdesitt}
\end{equation}
By feeding the recursion relation to a computer, and plotting the
trajectories of the folds, the minimal surface is constructed and seen to
have the properties described above, as depicted in Fig.6 } }

\section{Comments and Conclusions}

{\ {\ We have solved generally the classical 2D string theory in any curved
space-time. All stringy solutions correspond to folded strings. All
solutions tend to the BBHP\ solutions in the asymptotically flat region of
the curved space-time. Therefore, the BBHP solutions of eq.(\ref{general})
serve to classify all the solutions for any curved space-time. In fact, the
sign patterns of the BBHP solutions provide the method for dividing the
world-sheet into patches, thus defining the lattices associated with the $%
A,B,C,D$ solutions, as explained in the paragraph following eq.(14). The
matching of boundaries for these functions gives the general solution in
curved space-time in the form of a ``transfer matrix''. Thus, }}lattices on
the world-sheet plus geometry in space-time lead to transfer matrices. This
seems to be a rich area to explore in more detail.

{\ {\ The method was explicitly applied to the yo-yo solution, and the
general yo-yo solution in any curved space-time was constructed.
Specializing further the metric, the transfer matrices were derived for a
black hole space-time and for a cosmological space-time. } }

{\ {\ The general physical motion of the string follows on the average a
geodesic of the massive particle, consistent with intuition. However, the
stringy behavior becomes evident in the vicinity of singularities where new
phenomena, such as tunneling (similar to diffraction), take place. } }

{\ {\ Given the fact that the string in 2D is quite non-trivial classically,
we expect that there is a consistent quantization procedure that includes
the non-trivial folded states. We have already outlined in the introduction
that in covariant quantization (as well as in the semiclassical quantization
of folds carried out by BBHP)\ the 2D string in flat space-time has indeed
extra states corresponding to folded strings. A similar covariant
quantization can be carried out for the 2D black hole string by using the
Kac-Moody current algebra formulation, and relaxing the $c=26$ condition
(i.e. $k<9/4$) to include the folded strings. What would also be interesting
is to find the correct formulation for interacting folded strings. The path
integral approach started in \cite{yankielowicz} seems to be promising, and
it may be possible to make faster progress by reformulating it in the
conformal gauge and relating it to our classical solutions\footnote{{\ Note
that the definition of fold in ref.\cite{yankielowicz} does not take into
account that the map from the world sheet to spacetime may be many to one
(i.e. a region maped to a segment), as explained in footnote 3. This feature
may be important in the formulation of folds and their interactions in the
path integral approach. In particular, the description of folds in the
conformal gauge, as in the present paper, may eventually prove to be a more
convenient mathematical formulation than the one used in \cite{yankielowicz}.%
}}}}.

{\ {\ Folded strings exist in higher dimensions as well. One can display the
general solution in flat space-time in the temporal gauge
\begin{equation}
\begin{array}{c}
x^0=p^0\tau ,\quad \vec x(\tau ,\sigma )=\vec x_L(\sigma ^{+})+\vec
x_R(\sigma ^{-}),\quad (\partial _{+}\vec x_L)^2=p_0^2=(\partial _{-}\vec
x_R)^2 \\
\\
\partial _{+}\vec x_L=p^0\left( \frac{2{\bf f}}{1+{\bf f}^2},\frac{1-{\bf f}%
^2}{1+{\bf f}^2}\varepsilon _L\right) ,\quad \partial _{-}\vec x_R=p^0\left(
\frac{2{\bf g}}{1+{\bf g}^2},\frac{1-{\bf g}^2}{1+{\bf g}^2}\varepsilon
_R\right)
\end{array}
\label{dfolded}
\end{equation}
where ${\bf f}(\sigma ^{+}){\bf ,\,\,g(\sigma ^{-})}$ are arbitrary periodic
vectors in $d-2$ dimensions, {\it which could be discontinuous}, and $%
\varepsilon _L(\sigma ^{+}),\,\varepsilon _R(\sigma ^{-})$ take the values $%
\pm 1$ in patches of the corresponding variables such that the sign patterns
repeat periodically (as in the 2D\ string). When ${\bf f,g}$ are both zero
the solution reduces to the 2 dimensional BBHP case. In general, the
presence of discontinuous $\varepsilon _{L,}\varepsilon _R,$ and the
discontinuities in ${\bf f}(\sigma ^{+}){\bf ,\,\,g(\sigma ^{-})}$ gives a
larger set of solutions, which include strings that are partially or fully
folded. Discontinuities are allowed since the differential equations are
first order in the derivatives }$\partial _{+}$ and $\partial _{-}.${\ }Such
solutions are usually missed in the lightcone gauge even in the flat
classical theory (therefore, the lightcone ``gauge'' is not really a gauge).
}

{\ The curved space-time analogs of such solutions in higher dimensions are
presently under investigation. We suspect that the inclusion of the quantum
states corresponding to such solutions may lead to a consistent quantum
theory in less than 26 dimensions. As already emphasized earlier in the
paper, the free string is perfectly consistent as a quantum theory for $c<26$%
, including the folded states. The interacting quantum string with folds
remains as an open possibility. }

\newpage\


\newcommand\putfig[3]{
   \vbox{
   \let\picnaturalsize=N
   \def\picsize{#3}
   \def\picfilename{#1}
   \ifx\nopictures Y\else{\ifx\epsfloaded Y\else\input epsf \fi
   \let\epsfloaded=Y
   \centerline{\ifx\picnaturalsize N\epsfxsize \picsize\fi
   \epsfbox{\picfilename}}}\fi
   \vspace{1.0cm}
   {\it #2}
   \vspace{1.5cm}
   }
}

\begin{figure}

\putfig{ 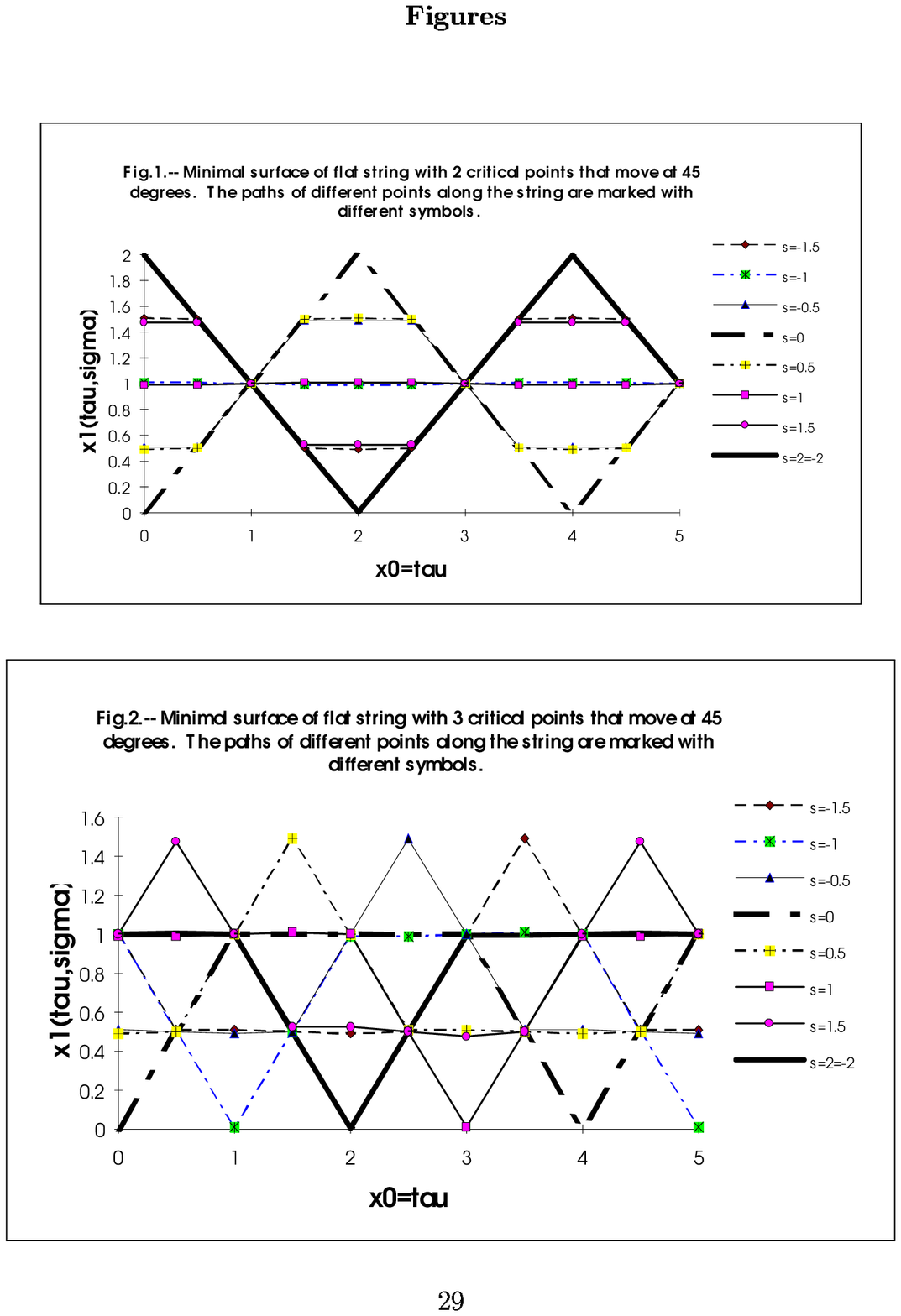}{150mm}{200mm}
\end{figure}

\begin{figure}
\putfig{ 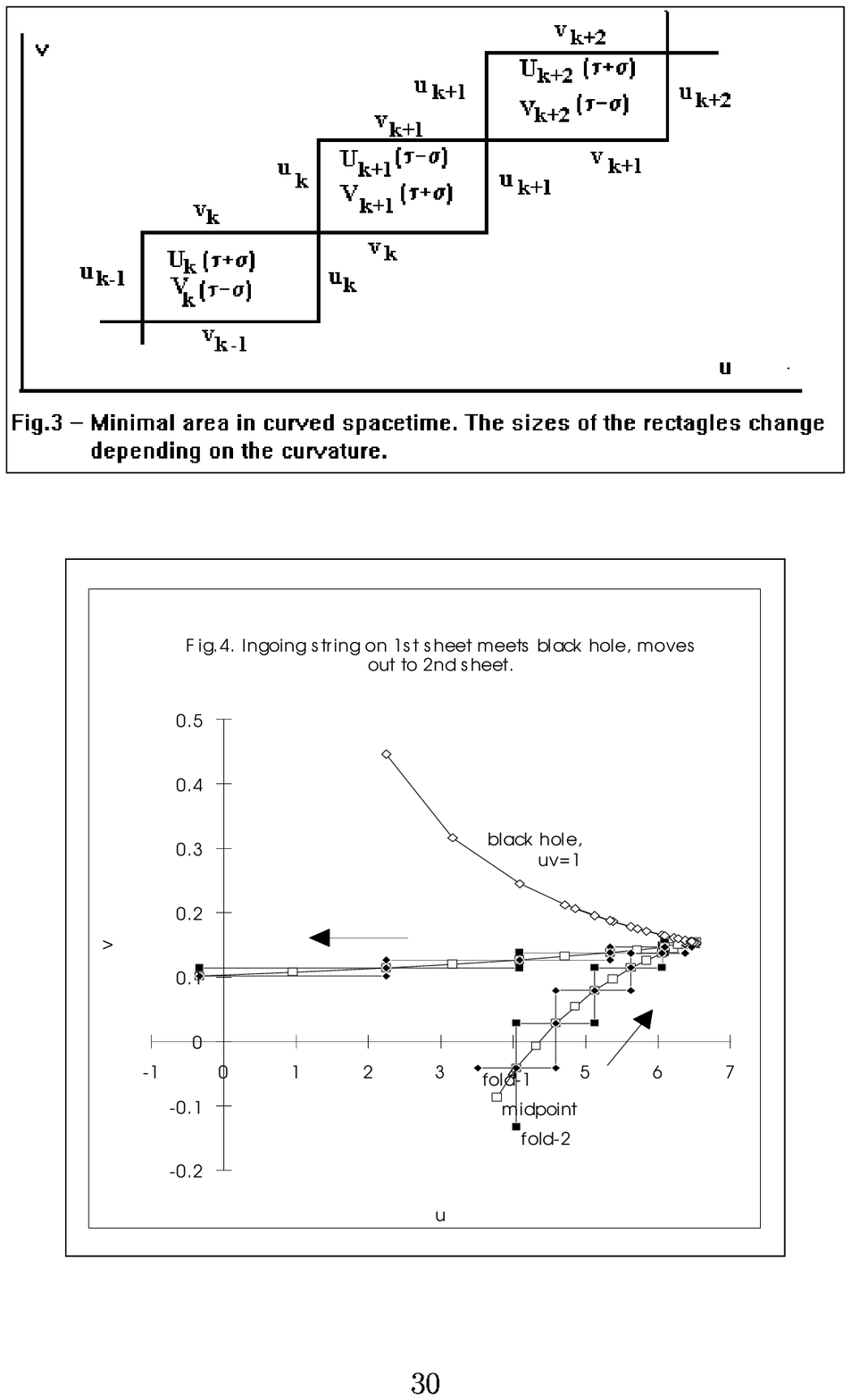}{150mm}{200mm}
\end{figure}

\begin{figure}
\putfig{ 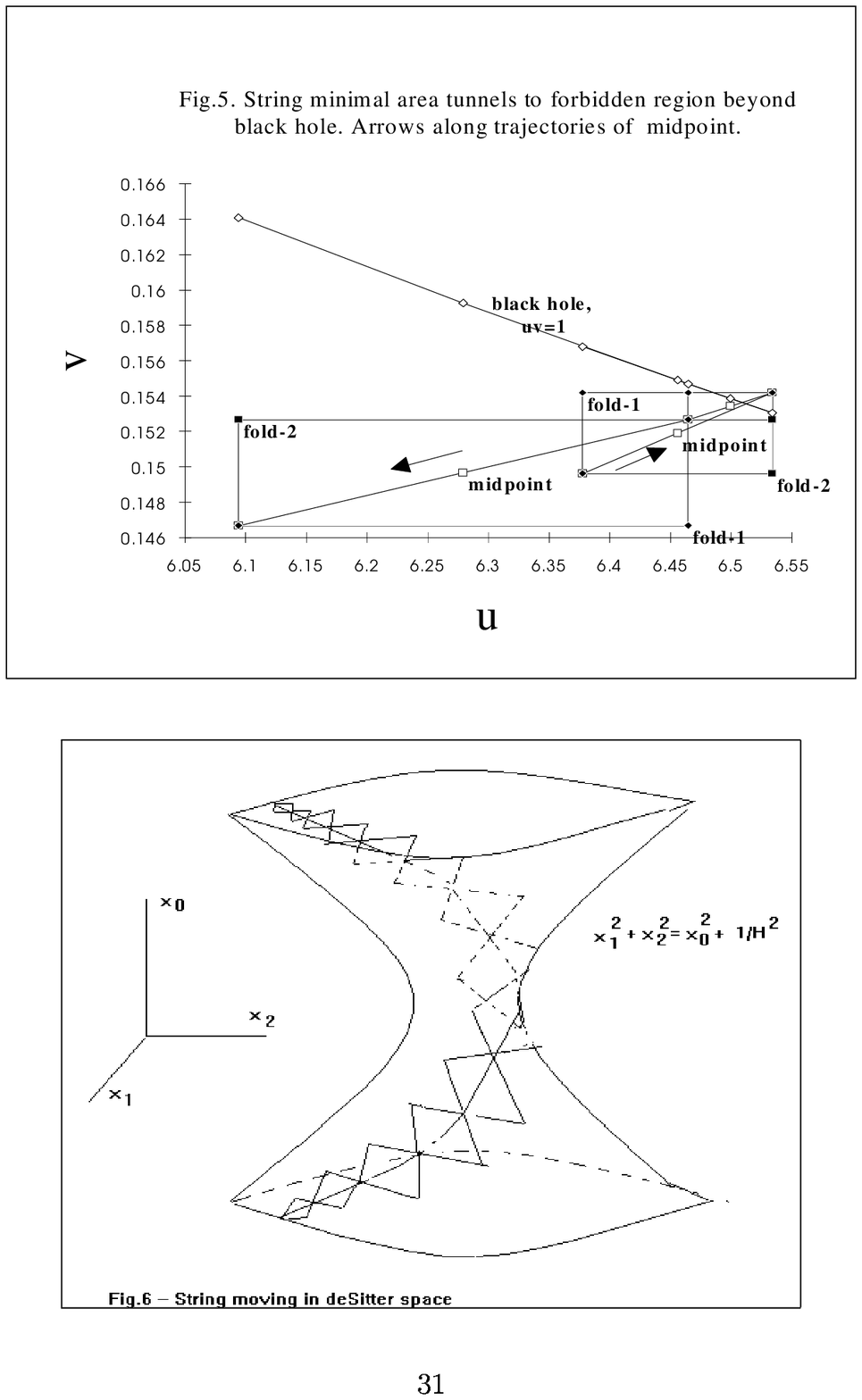}{150mm}{200mm}

\end{figure}
\end{document}
